\documentclass[journal]{IEEEtran}

\usepackage[compress]{cite}
\usepackage{stmaryrd}
\usepackage{mathrsfs}
\usepackage{amsfonts}
\usepackage{amsmath}
\usepackage{cases}
\usepackage{txfonts}
\usepackage{amssymb}
\usepackage{epsfig}
\usepackage{pdfpages}
\usepackage{algorithm}
\usepackage{algpseudocode}
\usepackage{slashbox}
\usepackage{multirow}
\usepackage{lipsum}
\usepackage{mathtools, cuted}

\newcommand{\fat}[1]{\mbox{\boldmath$#1$}}

\newtheorem{lemma}{Lemma}
\newtheorem{definition}{Definition}

\newcommand{\myQED}{\mbox{}\hfill{$\Box$}}

\begin{document}

\title{Near-Optimal Detection for Both Data and Sneak-Path Interference in Resistive Memories with Random Cell Selector Failures}

\author{Guanghui Song, \IEEEmembership{Member, IEEE},
        Kui Cai, \IEEEmembership{Senior Member, IEEE}, Ce Sun, \\ Xingwei Zhong, and Jun Cheng, \IEEEmembership{Member, IEEE}
}

\maketitle

\begin{abstract}
Resistive random-access memory is one of the most promising candidates for the next generation of non-volatile memory technology. However, its crossbar structure causes severe ``sneak-path" interference, which also leads to strong inter-cell correlation. Recent works have mainly focused on sub-optimal data detection schemes by ignoring inter-cell correlation and treating sneak-path interference as independent noise. We propose a near-optimal data detection scheme that can approach the performance bound of the optimal detection scheme. Our detection scheme leverages a joint data and sneak-path interference recovery and can use all inter-cell correlations. The scheme is appropriate for data detection of large memory arrays with only linear operation complexity. \end{abstract}


\section{Introduction}
Resistive random-access memory (ReRAM) is an emerging non-volatile memory technology that stores data over memristors. A memristor, referred to as a memory cell, has two states, the High-Resistance State (HRS) and the Low-Resistance State (LRS), reflecting the two logical values of a bit, logic 0 and logic 1. Many of the advantages of ReRAM as compared to conventional memory, such as ultra-dense data storage and parallel reading and writing, are derived from ReRAM's crossbar structure, in which the memory cells are arranged in a matrix array \cite{Strukov}. However, this crossbar structure also causes a severe interference problem called a ``sneak path" (SP) \cite{Zidan,Yuval}. Specifically, when a memory cell is read, voltage is applied to the memristor, and the resistance value is detected by sensing the current flowing through the memristor. The SPs are undesirable paths in parallel to the target cell that will decrease the sensed resistance value and make the cell more vulnerable to noise.

A common method to mitigate SPs is to introduce selectors in series with the memory cells. If all the selectors are functioning, the SPs can be eliminated. However, the selectors are also prone to failures due to imperfections in the memory fabrication and maintenance
process, leading to reoccurrence of the SPs.
Recently, Ben-Hur and Cassuto, Chen et al., and Song et al. \cite{Ben,CZH,SongArchive} addressed the data detection problem for ReRAM with random selector failures (SFs) using information and coding theories. An essential characteristic of the data detection in ReRAM is that the SP interference for cells within the same array is correlated, and there have been very few theories developed for data detection over correlated channels. A basic approach used by Ben-Hur and Cassuto, and Song et al. \cite{Ben,SongArchive} is to treat the SP interference as noise and detect each cell independently. Chen et al. \cite{CZH} developed a pilot assisted adaptive thresholding scheme that can use part of the inter-cell correlation during data detection. However, the pilot dispatching approach reduces the data rate, and when the array size is large the probability calculation used in Chen et al. \cite{CZH} becomes intractable. None of the above works guarantee near-optimal data detection performance, which has been unknown so far.

We propose a near-optimal data detection scheme for ReRAM with random SFs.
The scheme leverages joint data and SP interference recovery and can use all inter-cell correlations.
The main idea is that we  first detect the SF locations and the rows and columns that contain SFs (SF rows and columns). The SF rows and columns determine the SP locations over the whole array.  With the SF rows and columns as side information, we can
detect the remaining data independently based on the maximum likelihood (ML) rule. We show that when the array is large, the detection of SF rows and columns is very accurate, and therefore, the overall detection approaches the ML detection. We derive a bit error rate (BER) lower bound for the optimal data detection, and our scheme can achieve this bound. In addition, our scheme does not reduce the data rate and can be applied for data detection of very large arrays with only linear operation complexity.

The rest of this paper is organized as follows. In Section~\ref{sec:model}, we present the SP model and the data detection problem.
The main idea of our detection scheme is described in Section~\ref{sec:scheme}. In Section~\ref{sec:property}, we explain some properties about SPs, which are the foundation of the detection of SF locations and SF rows and columns. Detection methods for SF locations and SF rows and columns are proposed in Section~\ref{sec:SF-detection}. In Section~\ref{sec:non-SF}, we present non-SF row and column detection. Detection performance bound and simulation results are given in Section~\ref{sec:simulation}. We conclude in Section~\ref{sec:conclusion}.

\section{Sneak-Path Model}\label{sec:model}
An  $N\times N$ resistive crossbar memory array $\mathcal{R}_e$ contains $N^2$ resistive memory cells, in which the cell that lies at the $m$-th row and $n$-th column
is called cell $(m,n), 1\leq m, n\leq N$. Each memory cell has two states, the HRS $R_0$ and the LRS $R_1$, which can represent two states of one bit, logical ``0" and logical ``1".
We refer to a cell with an HRS/LRS as an HRS/LRS cell.
Therefore, the  $N\times N$ crossbar memory array can store an $N\times N$ binary data array $\fat{X}=[x_{m,n}]_{N\times N}, x_{m,n}\in\{0, 1\}$. We assume that the entries of $\fat{X}$ are with i.i.d. Bernoulli $(q)$, i.e., $\textrm{Pr}(x_{m,n}=1)=q$ and $\textrm{Pr}(x_{m,n}=0)=1-q$ for $0<q<1$ and $m, n=1,..., N$.

\begin{figure}[t]
\includegraphics[width=
2.2 in]{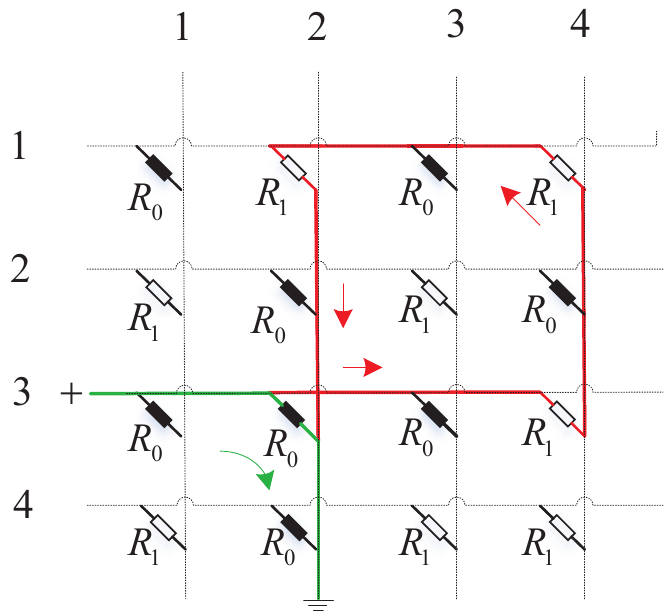}
\centering
\caption{Example of an SP during the reading of cell $(3, 2)$ in a $4\times 4$ memory array. The green line is the desired path for resistance measuring. $(3, 2)\rightarrow(3, 4)\rightarrow(1, 4)\rightarrow(1, 2)\rightarrow(3, 2)$ forms an SP (red line) in parallel to target cell $(3, 2)$ that degrades the measured resistance value. Note that the horizontal and vertical lines are connected via intersectional memory cells. Arrows show current flow directions. A reverse current flows across cell $(1, 4)$.} \label{fig:SPmodel}
\end{figure}

To read data array $\fat{X}$ from the memory array, we have to detect the resistance state of each memory cell. In general, if a cell is detected as HRS, the corresponding bit is identified as a ``0"; if it is detected as LRS, the bit is identified as a ``1". The major challenge is the existence
of SP interference, which can distort the detected resistance value. Specifically, when a cell is read, a voltage is applied to the target cell to measure its resistance. AN SP is an undesirable path in parallel to the target cell that will affect the resistance detection. Fig.~\ref{fig:SPmodel} shows an example of a crossbar array with storage for data array
\begin{equation}
\fat{X}=\begin{bmatrix}
0 & 1 & 0 & 1\\
1 & 0 & 1 & 0\\
0 & 0 & 0 & 1\\
1 & 0 & 1 & 1
\end{bmatrix}.
\end{equation}
 Here $(3, 2)$ is a target cell for reading, and the green line is the desired current path for resistance measuring. However, when $(3, 4), (1, 4), and (1, 2)$ are all LRS cells, an SP forms, as indicated by the red line. A direct impact of the SP is a reduction of the readout signal. Thus, the detected resistance value in this case becomes
\begin{equation}
R_0^\prime=\left(\frac{1}{R_0}+\frac{1}{R_s}\right)^{-1}<R_0
\end{equation}
where $R_s$ is the parasitic resistance value brought by the SP. In general, an SP is defined as a path that originates from and returns to an HRS cell while traversing 3 LRS cells through alternating vertical and horizontal steps. Note that SPs benefit the detection when the target cell is an LRS cell. They are detrimental only when an HRS cell is read, making it more vulnerable to noise. For this reason, we only consider SPs here when an HRS cell is read.

The most widely-used method to mitigate SP interference is to introduce a cell selector in series to each memory cell. A cell selector is an electrical device that allows current to flow only in one direction across the cell. Since an SP inherently produces a reverse current in at least one cell along the parallel path (eg. cell $(1, 4)$ in Fig.~\ref{fig:SPmodel}), if all the cell selectors work they can completely eliminate SPs from the entire array.
 However, selectors are also prone to failures due to imperfections in production or maintenance of a memory array, leading to reoccurrence of the SPs. In the case of Fig.~\ref{fig:SPmodel}, due to the circuit structure of the crossbar array, cells $(3, 4)$ and $(1, 2)$ will conduct current in the forward direction and not be affected by their selectors. Only when the selector of diagonal cell $(1, 4)$ is faulty will an SP be formed.
Following Ben-Hur and Cassuto, as well as Chen et al. \cite{Ben,CZH}, we assume selectors fail randomly, and the SF pattern is unknown to the data detector.

If a cell is affected by an SP during its reading, we call it an SP cell, otherwise, we call it a non-SP cell. Therefore, the memory array may contain three types of cells: HRS cells, LRS cells, and SP cells.
Generally, cell $(m, n)$ is an SP cell if the following conditions are satisfied:

\emph{[Conditions of $(m, n)$ Being an SP Cell:]}

1) $x_{m, n}=0$.

2) We can find at least one combination of $i, j\in\{1,...,N\}$ in $\fat{X}$ that satisfies
\begin{equation}
x_{m, j}=x_{i, j}=x_{i, n}=1.
\end{equation}

3) The selector at diagonal cell $(i, j)$ fails.

These SP cell conditions limit the SPs to a length of 3, i.e., traversing three cells. Following Ben-Hur and Cassuto \cite{Ben} and Chen et al. \cite{CZH}, we neglect the affection of longer SPs since compared to the length-3 SPs their affection is insignificant.
Also following Chen et al. \cite{CZH}, we do not consider the superposition effect of multiple SPs to one cell. More sophisticated SP models were considered in Ben-Hur and Cassuto \cite{Ben}, and the principle of our work can be extended to those models.

 To formulate the readout signal, we first define an SP cell indicator $e_{m, n}$ for cell $(m, n)$ to be a Boolean variable with the value $e_{m, n}=1$ if $(m, n)$ is an SP cell, otherwise, $e_{m, n}=0$ if $(m, n)$ is a non-SP cell.

The readback signal array $\fat{Y}=[y_{m, n}]_{N\times N}$ is given by:
\begin{eqnarray}
\!\!\!\!\!\!\!\!\!\!\!&&y_{m, n}=r_{m,n}+\eta_{m, n}\label{eq:readout}\\
\!\!\!\!\!\!\!\!\!\!\!&&\textrm{with}\ \ r_{m, n}=\begin{cases}\left( (\frac{1}{R_0}+\frac{e_{m, n}}{R_s}\right)^{-1}&\mbox{if $x_{m, n}=0$}\\ R_1&\mbox{if $x_{m, n}=1$} \end{cases}
\end{eqnarray}
where $\eta_{m, n}\sim \mathcal{N}(0, \sigma^2), m=1,...,N, n=1,...,N$, is an additive Gaussian noise with mean 0 and variance $\sigma^2$ \cite{Ben}. We use this Gaussian noise model, following Ben-Hur and Cassuto \cite{Ben} and Chen et al. \cite{CZH}, to model a mix of various noises of the ReRAM system.  Note that there are noise sources that are not Gaussian \cite{Yu2012, Wong2012}, while our work can be easily extended to other distributions of noise.

\begin{figure}[t]
\includegraphics[width=
3.5 in]{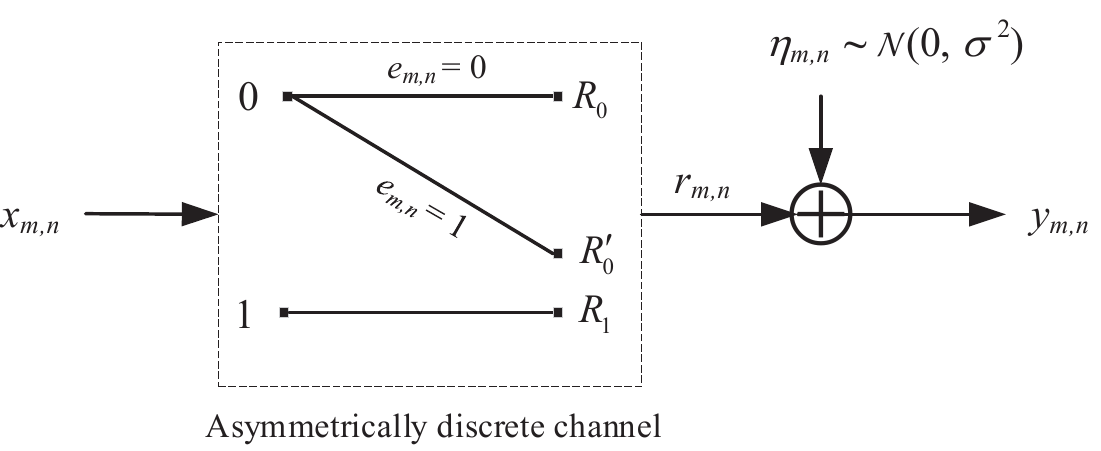}
\centering
\caption{ReRAM channel model} \label{fig:channel}
\end{figure}

We call $\fat{R}=[r_{m,n}]_{N\times N}$ the readout resistance array. Fig.~\ref{fig:channel} illustrates the ReRAM channel model, which is a concatenation of an asymmetrical discrete channel and an additive Gaussian noise channel.

\section{Detection Scheme}\label{sec:scheme}

The fundamental problem of ReRAM data detection is to recover stored data array $\fat{X}$ based on readback signal array $\fat{Y}$ in the presence of SP interference $[e_{m, n}]_{N\times N}$ and Gaussian noise $[\eta_{m, n}]_{N\times N}$.
 In this section, we propose our detection scheme.

    For a given resistive crossbar array $\mathcal{R}_e$, let
    \begin{equation}
    \varphi_{sf}=\left\{(i, j)|\  \textrm{selector fails at}\ (i, j)\right\}
     \end{equation}
     be the set that includes all the indices of SFs. After writing data array $\fat{X}$ into $\mathcal{R}_e$, let $\varphi_{sf}^*=\{(i, j)| (i, j)\in \varphi_{sf}, x_{i, j}=1\}$ be the set that includes all the indices with both SFs and logical-1 storage at the corresponding memory cell. A SF whose index is in $\varphi_{sf}^*$ is called an active SF. Based on the SP cell condition, only positive SFs can cause SPs.

     \begin{definition}
     If $(i, j)\in\varphi_{sf}^*$, row $i$ of $\fat{X}$ is called an SF row and column $j$ is called an SF column. If a row or column is not an SF row or column, we call it a non-SF row or column.
     \end{definition}

     We first need to reformulate the readout resistance array. Let
  \begin{equation}
  R_x(e)=\left(\frac{1}{R_x}+(1-x)\frac{e}{R_s}\right)^{-1},\ \ \textrm{for} \ x, e \in\{0, 1\}.
   \end{equation}
   Thus, $R_1(0)=R_1(1)=R_1$ and $R_0(0)=R_0, R_0(1)=R_0^\prime$.
     Based on the SP occurrence condition, the readout resistance in (\ref{eq:readout}) can be rewritten as
\begin{eqnarray}
r_{m,n}\!\!\!\!\!\!\!&&= R_{x_{m,n}}\left(e_{m,n}\right)\\
&&= R_{x_{m,n}}\left(\bigcup_{(i,j)\in\varphi_{sf}}x_{i,n}x_{i,j}x_{m,j}\right)\\
&&= R_{x_{m,n}}\left(\bigcup_{(i,j)\in\varphi_{sf}^*}x_{i,n}x_{m,j}\right)\label{eq:readout2}
\end{eqnarray}
where $\bigcup$ is the logical OR operator, i.e., $\bigcup_{(i,j)\in\varphi_{sf}}x_{i,n}x_{i,j}x_{m,j}=1$ if at least one of $(i,j)\in\varphi_{sf}$ with $x_{i,n}x_{i,j}x_{m,j}=1$ exists, otherwise, $\bigcup_{(i,j)\in\varphi_{sf}}x_{i,n}x_{i,j}x_{m,j}=0$. Therefore, the readout resistance of cell $(m, n)$ is a function of both $x_{m, n}$ and the SF rows and columns.

\begin{figure}[t]
\includegraphics[width=
2.5 in]{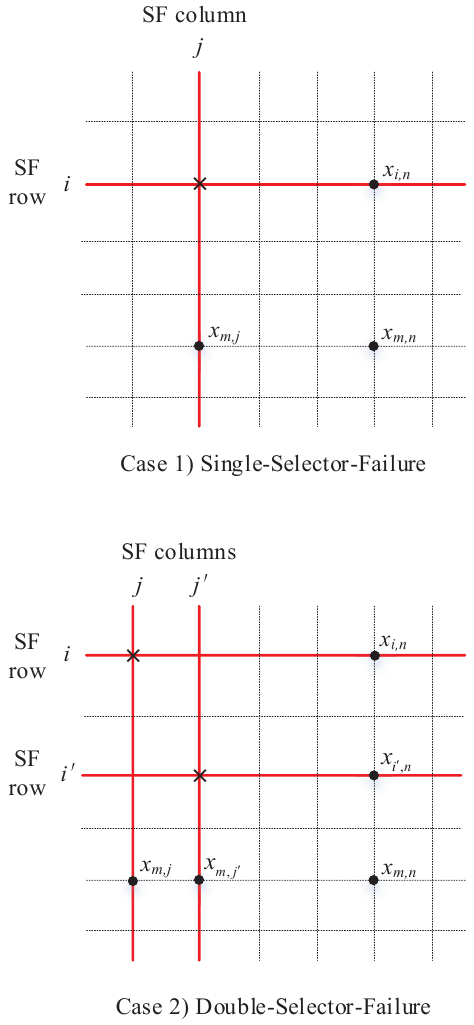}
\centering
\caption{Illustration of memory arrays of Case 1) Single-Selector Failure and Case 2) Double-Selector Failure, where $\times$ is labeled at the cell with SFs.} \label{fig:roudout-signal}
\end{figure}

 Since only the active SFs affect the data detection performance, in the rest of this paper any SF that we refer to means active SF.
Since, as considered in Cassuto et al. and Chen et al. \cite{CZH,Yuval}, a selector fails with very low probability, the most likely situation is one in which only very few selectors fail in an array. In this paper, we assume the number of SFs in a single resistive array is less than three. In addition, we assume two SFs do not lie in the same row or column, which is also the most likely case. Therefore, each resistive memory array can have three possible cases: Case 0) no  SF $\varphi^*=\emptyset$; Case 1) Single-Selector Failure $\varphi^*=\{(i, j)\}$; Case 2) Double-Selector Failure $\varphi^*=\{(i, j), (i^\prime, j^\prime)\}, i\neq i^\prime, j\neq j^\prime$.  In Case 0), no SP occurs. In Cases 1) and 2), SPs may occur.  Memory arrays of Cases 1) and 2) are shown in Fig.~\ref{fig:roudout-signal}.
 Their readout resistance at cell $(m, n)$ are\\ \vspace{2mm}
\emph{ [Case 1) Single-Selector Failure at (i, j):]}
\begin{eqnarray}
r_{m,n}=\begin{cases} R_0&\mbox{if $x_{m, n}=0,\ x_{i,n}x_{m,j}=0$}\\  R_0^\prime&\mbox{if $x_{m, n}=0, \ x_{i,n}x_{m,j}= 1$}\\ R_1 &\mbox{if $x_{m, n}=1$} \end{cases}
\label{eq:readout-eg1}
\end{eqnarray}\\ \vspace{2mm}
\emph{ [Case 2) Double-Selector Failure at $(i, j), (i^\prime, j^\prime)$:]}
\begin{eqnarray}
r_{m,n}=\begin{cases} R_0&\mbox{if $x_{m, n}=0,\ x_{i,n}x_{m,j}= x_{i^\prime,n}x_{m,j^\prime}=0$}\\  R_0^\prime&\mbox{if $x_{m, n}=0, \ x_{i,n}x_{m,j}= 1\ \textrm{or}\ x_{i^\prime,n}x_{m,j^\prime}=1$}\\ R_1 &\mbox{if $x_{m, n}=1$}. \end{cases}
\label{eq:readout-eg2}
\end{eqnarray}

\begin{figure}[t]
\includegraphics[width=
3.0 in]{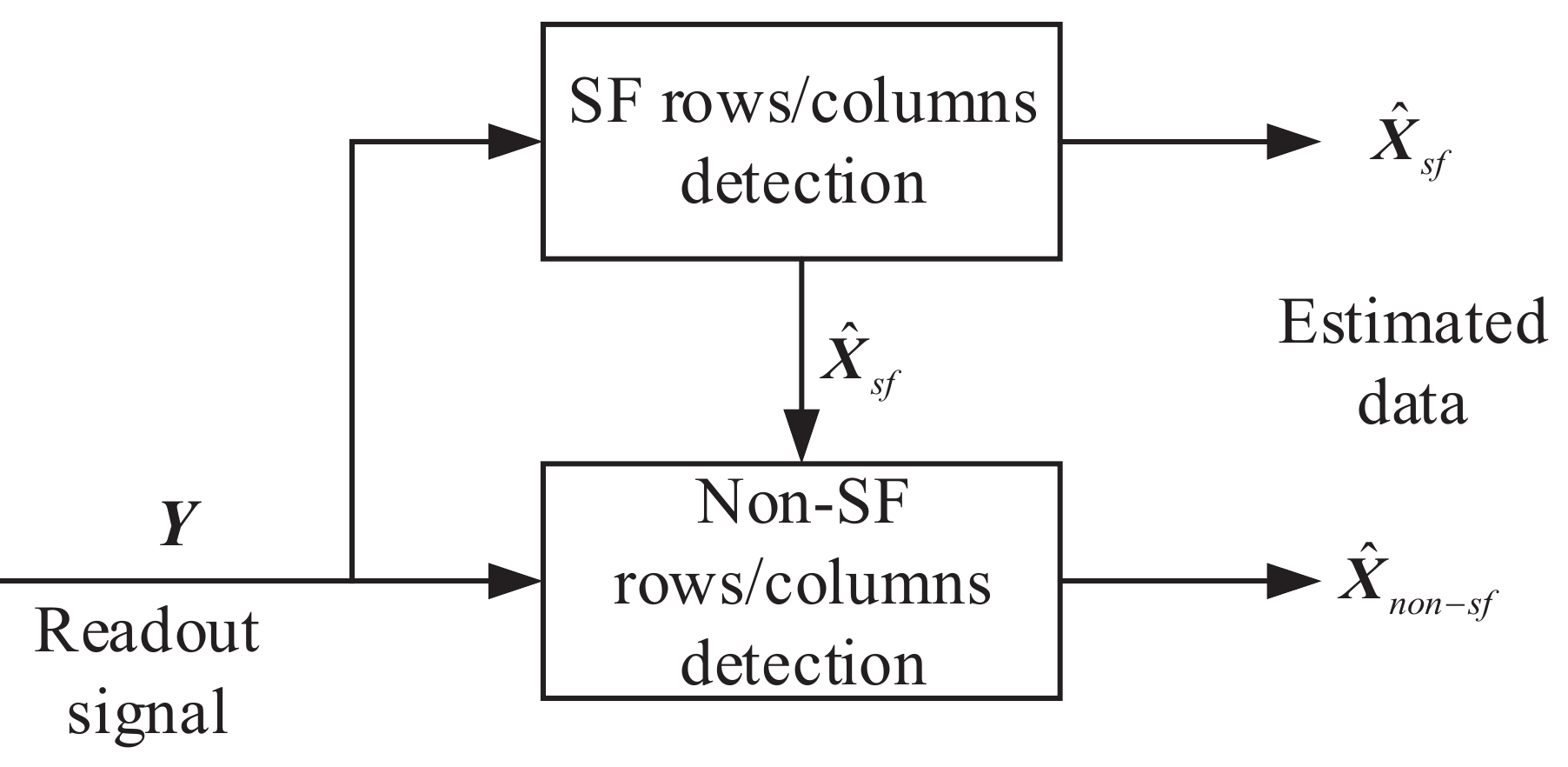}
\centering
\caption{Sketch of detection scheme} \label{fig:detection}
\end{figure}

 Since all the readout signals in the array are correlated, all correlated to the SF rows and columns, when the array size is large, MAP detection $\max_{\fat{X}}\sum_{\varphi^*}\textrm{Pr}(\varphi^*)\textrm{Pr}\left(\fat{X}|\fat{Y}, \varphi^*\right)$ which requires globally maximizing the posteriori probability by taking account of all the possible data patterns and SF patterns is impossible. Once the SF rows and columns are known, the remaining data can be detected based on a bit-wise MAP rule, leading to the following detection scheme. We split data array $\fat{X}$ into two parts, $\fat{X}_{sf}$ and $\fat{X}_{non-sf}$, where $\fat{X}_{sf}$ contains the entries of SF rows and columns and $\fat{X}_{non-sf}$ contains all the remaining entries.
Since all the readout signals are related to the SF rows and columns, we first detect $\fat{X}_{sf}$ based on the whole readout signal array $\fat{Y}$ by regarding $\fat{X}_{non-sf}$ as random variables. Based on $\hat{\fat{X}}_{sf}$, we can detect $\fat{X}_{non-sf}$ with the MAP criterion (a sketch of the detection is shown in Fig.~\ref{fig:detection}). We will show that when the SF number is not more than three, the detection of $\fat{X}_{sf}$ is quite accurate. Therefore, the overall detection performance will approach that of the MAP detection.

\section{Sneak-Path Distribution Properties}\label{sec:property}
We discuss some SP distribution properties, which are the theoretical foundation for SF row and column detection.

Before we discuss these properties, we need to clarify that when we refer to an array, it can have two interpretations, resistance array $\fat{R}$ with cells and resistive states or
associated data array $\fat{X}$ with bit values. In the following, the array we refer to is a combination of $\fat{R}$ and $\fat{X}$, and it has both resistive states and bit values. In different situations, we use different aspects of the array.

\begin{definition}
A row or column is called an SP row or column if it contains SP cells, and it is called a non-SP row or column if it does not contain an SP cell.
\end{definition}

Assume there is a SF at cell $(i, j)$. The corresponding SF row and column are $(x_{i,1}, x_{i,2}, ..., x_{i,N})$ and $(x_{1,j}, x_{2,j}, ..., x_{N,j})$. We have the following definition.
\begin{definition}
Location $(i, j)$ is called the SF center. Each non-zero entry of $x_{i,n}=1$ and $x_{m, j}=1$ for $n\neq j$ and $m\neq i$ in the SF row or column is called an SP support.
\end{definition}

SF row $i$, SF column $j$, and SP supports $x_{i,n}=1, x_{m, j}=1$ for $n\neq j, m\neq i$ are caused by the SF at cell $(i, j)$.


Although an SF row or column may contain multiple SP supports, a non-SF row or column usually contain very few SP supports (only at the intersections with the SF columns and rows).
Since we assume that the maximum SF number is two in an array, a non-SF row or column can only contain at most two SP supports.

\begin{definition}
A row or column is called an SP-supported row or column if it contains an SP support; it is
called a single-SP-supported row or column if it contains only one SP support; and it is called a double-SP-supported row or column if it contains two SP supports.
\end{definition}

 If a non-SF row or column is a double-SP-supported row or column, it should contain two SP supports that are caused by two different SFs.

The following lemmas  are directly from the SP cell occurrence condition.
\begin{lemma}
 An SP row or column must be an SP-supported row or column.
\end{lemma}

\begin{lemma}\label{lem:1}
Cell $(m, n)$ with $x_{m,n}=0$ is an SP cell if and only if there exists an SP support in row $m$ and an SP support in column $n$ that are caused by the same SF.
\end{lemma}

\begin{definition}
A cell at the intersection of an SP-supported row and an SP-supported column is called a critical cell.
\end{definition}

Critical cells play a very important role in our data detection. There are three possible states for a critical cell corresponding to three situations for the data storage, i.e.,

(1) LRS: if and only if the stored data is a logical-1;

(2) HRS: if and only if the stored data is a logical-0 and the cell is at the intersection of an SP-supported row and column that are caused by different SFs; or,

(3) SP: if and only if the stored data is a logical-0 and the cell is at the intersection of an SP-supported row and column that are caused by the same SF.

\begin{definition}\label{def:6}
An SP row or column whose critical cells are either LRS cells or SP cells is called a complete-SP row or column; otherwise, if any of its critical cells is an HRS cell, it is called an incomplete-SP row or column.
\end{definition}

\begin{figure*}[t]
\includegraphics[width=
4.0 in]{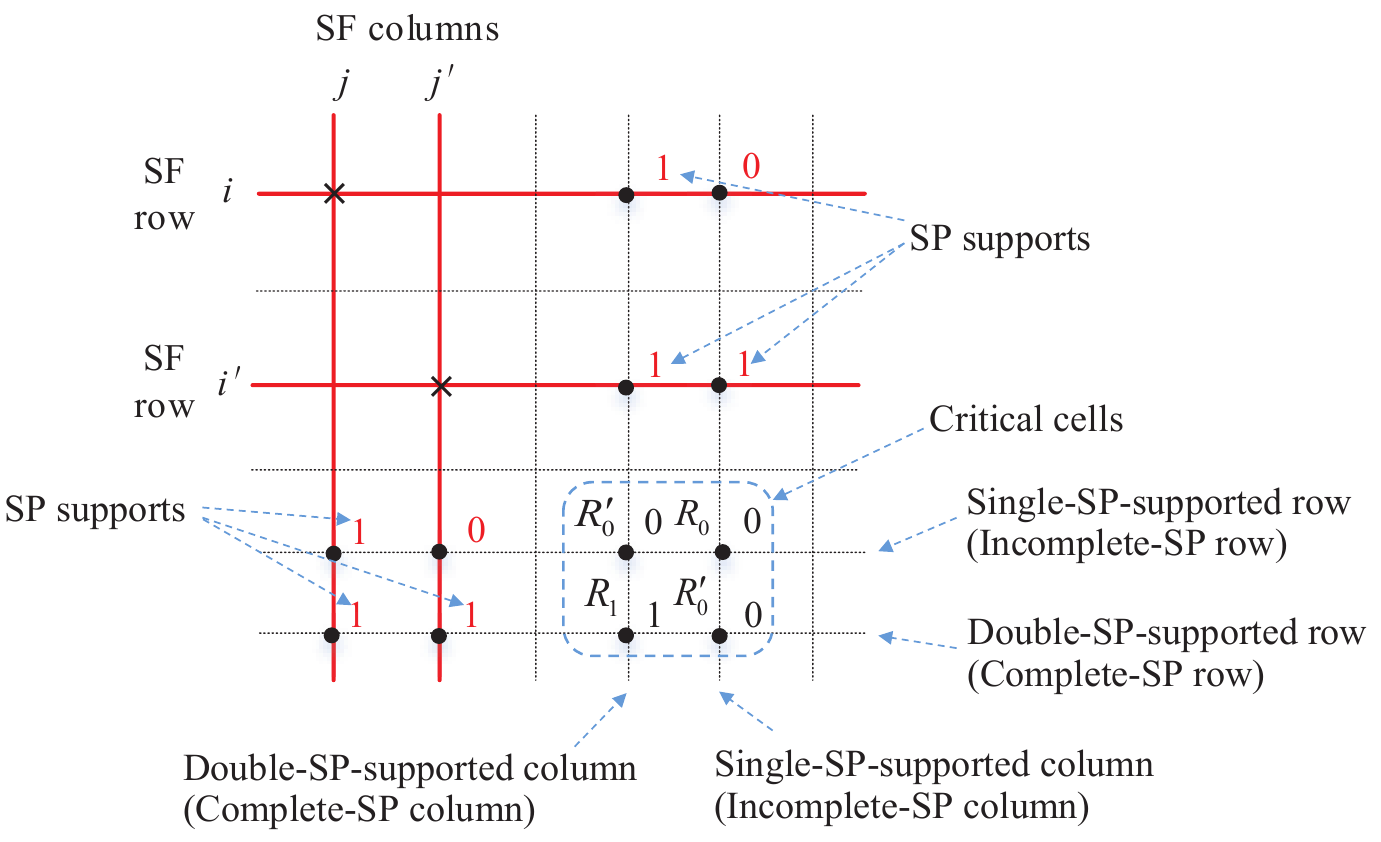}
\centering
\caption{An illustration for SP-supported rows and columns, critical cells, and different types of SP rows and columns} \label{fig:Example-definitions}
\end{figure*}

Figure~\ref{fig:Example-definitions} illustrates some examples of SP-supported rows and columns, critical cells, and different types of SP-rows and columns. The figure shows an array with double-SFs. There are two SP-supported rows and two SP-supported columns whose intersectional cells are four critical cells. The critical cell
at the intersection of the single-SP-supported row and the single-SP-supported column is an HRS cell rather than an SP cell although the stored data is logical-$0$. This is because the cell belongs to an SP-supported row and column that are caused by two different SFs, i.e., $(i, j)$ and $(i^\prime, j^\prime)$. The critical cell
at the intersection of the double-SP-supported row and the double-SP-supported column is an LRS cell since the stored data is logical-$1$.
The other two critical cells are SP cells. Since the single-SP-supported row and column contains a critical cell with HRS, it is an incomplete-SP row and column.

Since a critical cell is an HRS cell only when the cell is at the intersection of an SP-supported row and column that are caused by different SFs, which does not happen in the Single-Selector Failure, we have the following lemma.

\begin{lemma}\label{lem:3}
 For Single-Selector Failure, all SP rows and columns are completed SP rows and columns.
\end{lemma}

It is easy to see that for Single-Selector Failure (\ref{eq:readout-eg1}), an SP does not happen at the SF row or column, which leads to the following lemma.
 \begin{lemma}\label{lem:SFSPtype}
 For Single-Selector Failure, the SF row or column is a non-SP row or column.
\end{lemma}

In Lemmas~\ref{lem:4}-\ref{lem:6}, we will discuss some relationships between the SP support numbers in a row or column and its SP-row or -column type.

\begin{lemma}\label{lem:4}
 For Single-Selector Failure, the probability of a non-SF and SP-supported row or column being a complete-SP row or column is $1-(1-(1-q)q)^{N-1}$.
\end{lemma}

\emph{Proof:} For Single-Selector Failure, a cell in a non-SF and SP-supported row or column is an SP cell if and only if the stored data at the cell is logical-0 and
there exists another SP support in the same column or row. This probability is $(1-q)q$. Therefore, the row or column is an SP row or column with probability $1-(1-(1-q)q)^{N-1}$ (at least one cell is an SP cell). Since for Single-Selector Failure all SP rows and columns are complete-SP rows and columns, we get the lemma.
 \myQED

\begin{lemma}\label{lem:5}
For Double-Selector Failure, a non-SF, double-SP-supported row or column is a complete-SP row or column with probability $1-(q+(1-q)^3)^{N-2}$, and
   a non-SF, complete-SP row or column is a double-SP-supported row or column with probability no less than $ 1-2\left(1-q(1-q)^2\right)^{N-2}/q^2$.
\end{lemma}

\emph{Proof:} Since for each critical cell in a non-SF, double-SP-supported row or column, we can always find an SP support in the same row and an SP support in the same column that are caused by the same SF,
 the cell can only be an LRS cell or an SP cell. Therefore, if a non-SF, double-SP-supported row or column is an SP row or column, it can only be a complete-SP row or column. Since a cell in a non-SF, double-SP-supported row or column is an SP cell with probability $(1-q)(1-(1-q)^2)=1-q-(1-q)^3$ (the stored data is a logical-0 and at least one SP support in the corresponding column or row), the row or column is an SP row or column with probability $1-(q+(1-q)^3)^{N-2}$ (at least one cell is an SP cell).

To prove the later part, we define the following events for a non-SF row or column: $E=\{\textrm{It is a complete-SP row or column}\}$, $E_k=\{\textrm{It contains}\ k\ \textrm{SP supports}\}, k=0, 1, 2$.
We want to calculate the probability
\begin{eqnarray}
\textrm{Pr}(E_2|E)\!\!\!\!\!\!\!\!&&=\frac{\textrm{Pr}(E_2)\textrm{Pr}(E|E_2)}{\textrm{Pr}(E)}\\
\!\!\!\!\!\!\!\!&&=\frac{\textrm{Pr}(E_2)\textrm{Pr}(E|E_2)}{\sum_{k=0}^2\textrm{Pr}(E_k)\textrm{Pr}(E|E_k)}.
\end{eqnarray}
 Since for Double-Selector Failure, $\textrm{Pr}(E_k)=\binom{2}{k}q^k(1-q)^{2-k}$, $\textrm{Pr}(E|E_0)=0$, and $\textrm{Pr}(E|E_2)=1-(q+(1-q)^3)^{N-2}$, we only need to calculate $\textrm{Pr}(E|E_1)$ to determine $\textrm{Pr}(E_2|E)$.
 If a row or column is a single-SP-supported row or column, it should contain only one SP support that is caused by one of the SFs. If it is also a complete-SP row or column, each of its cells can be an LRS cell with probability $q$, otherwise, it can
 be an SP cell with probability $(1-q)q$ (an SP support caused by the same SF exists at the same column or row), or a non-critical cell with probability $(1-q)^3$ (no SP support at the same column or row). Thus, we have
 $\textrm{Pr}(E|E_1)\leq(2q-q^2+(1-q)^3)^{N-2}=(1-q(1-q)^2)^{N-2}$ leading to
 \begin{eqnarray}
\textrm{Pr}(E_2|E) \!\!\!\!\!\!\!\!\!\!&&\geq 1-\frac{\textrm{Pr}(E_1)\textrm{Pr}(E|E_1)}{\textrm{Pr}(E_2)\textrm{Pr}(E|E_2)}\\ \!\!\!\!\!\!\!\!\!\!&&\geq1-\frac{2(1-q)\left(1-q(1-q)^2\right)^{N-2}}{q\left(1-(q+(1-q)^3)^{N-2}\right)}\\
\!\!\!\!\!\!\!\!\!\!&&\geq 1-2\left(1-q(1-q)^2\right)^{N-2}/q^2 \label{eq:unequal}
\end{eqnarray}
where we used $1-(q+(1-q)^3)^{N-2}\geq q(1-q)$ to get (\ref{eq:unequal}).
 \myQED

\begin{lemma}\label{lem:6}
 For Double-Selector Failure, each incomplete-SP row or column is a single-SP-supported row or column. A single-SP-supported row or column is an incomplete-SP row or column with probability no less than $1-2(1-q(1-q)^2)^{N-2}$.
\end{lemma}

\textit{Proof:} The former part is because an SP row or column must be an SP-supported row or column, and
 a double-SP-supported row or column can only be a non-SP row or column or a complete-SP row or column.

To prove the later part, we first have that for Double-Selector Failure, a single-SP-supported row or column is a non-SP row or column with probability no larger than $(q+(1-q)^2)^{N-2}$, and a completed-SP row or column with probability no larger than $(1-q(1-q)^2)^{N-2}$ (shown in the proof of Lemma~\ref{lem:5}). Since $q+(1-q)^2\leq 1-q(1-q)^2$, we get the lemma.
\myQED

In Lemmas~\ref{lem:4}-\ref{lem:6}, we showed that if a row or column contains a certain number of SP supports, it will be a certain type of  SP row or column or be a certain type of SP row or column with a probability no less than $1-f(q, N)$, where $\lim_{N\rightarrow\infty} f(q, N)= 0$, and we also showed that the reverse also holds. Therefore, we demonstrated a one-to-one correspondence between the number of SP supports in a row or column and its asymptotical SP-row or -column type as $N\rightarrow\infty$.
This one-to-one correspondence is summarized in TABLE~\ref{tab:SP-type}.

\begin{table}[t]
\caption{One-to-one correspondence between SP support numbers and asymptotical SP-row or -column types as $N\rightarrow\infty$}
\label{tab:SP-type}
\begin{center}
\begin{tabular}{|c|c|c|}
\hline\hline
SF type&SP support number& SP-row or -column type\\
\hline
\multirow{2}{*}{Single}&0& Non-SP\\
&1& Complete-SP\\
\hline
\multirow{3}{*}{Double}&0& Non-SP\\
&1& Incomplete-SP\\
&2& Complete-SP\\
\hline\hline
\end{tabular}
\end{center}
\end{table}

 \begin{figure*}[t]
\includegraphics[width=
5.0 in]{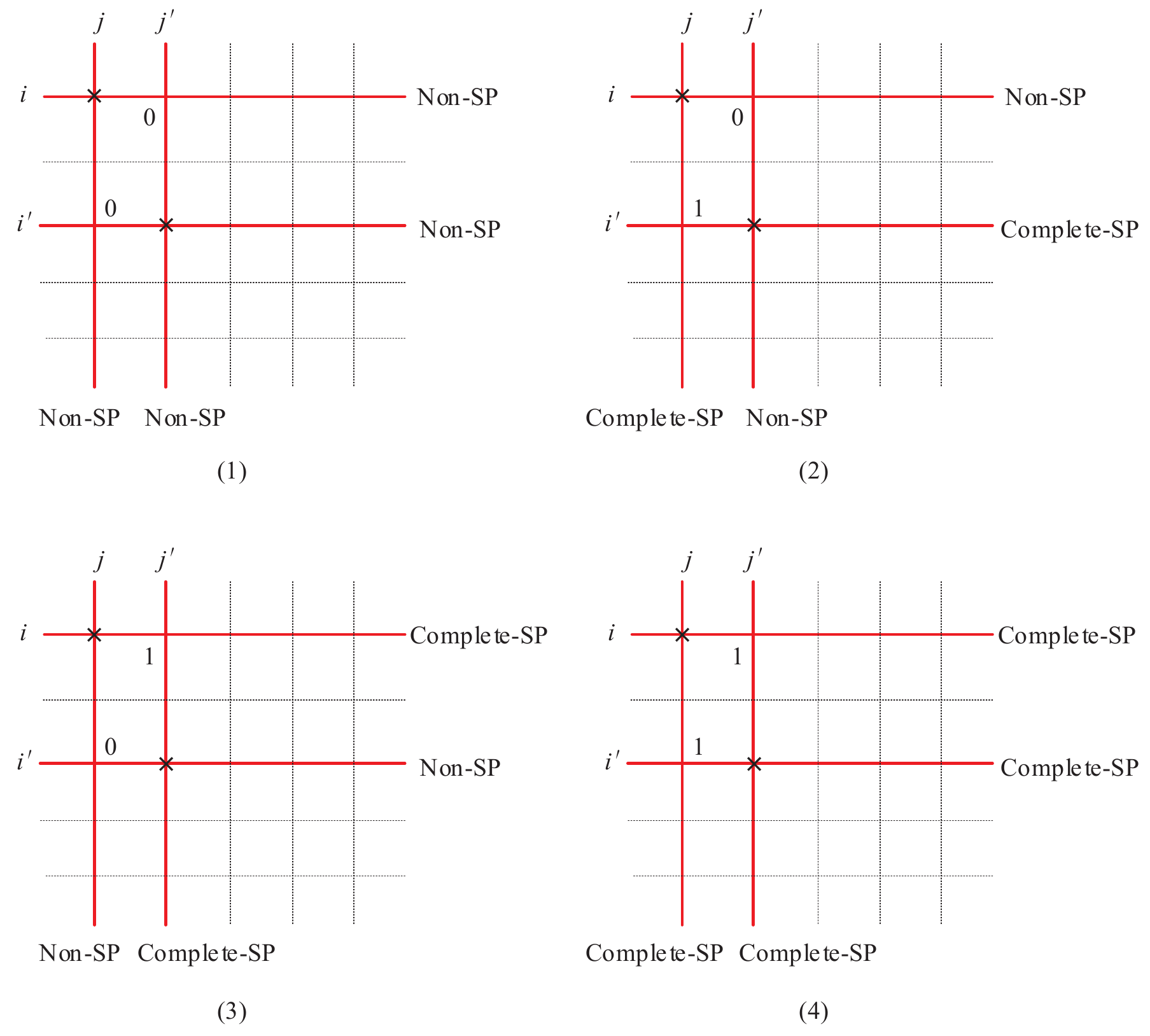}
\centering
\caption{One-to-one correspondence between SF row and column intersectional entries and their asymptotical SP-row or -column types as $N\rightarrow\infty$.} \label{fig:SFrowSP}
\end{figure*}

The following lemma reveals the SP-row or -column types of the SF rows or columns for Double-Selector Failure.
 \begin{lemma}\label{lem:7}
 For Double-Selector Failure with SFs at cells $(i, j)$ and $(i^\prime, j^\prime)$, if $x_{i,j^\prime}=0$, SF row $i$ and SF column $j^\prime$ are a non-SP row and column, respectively.
 If $x_{i,j^\prime}=1$, SF row $i$ or column $j^\prime$ are a complete-SP row or column with probability $1-(q+(1-q)^2)^{N-2}$. Similar conclusions hold for SF row $i^\prime$ and SF column $j$.
\end{lemma}

\emph{Proof:} To prove the former part, we notice that if $x_{i,j^\prime}=0$, according to (\ref{eq:readout-eg2}), the readout resistance $r_{i, n}=R_0$ or $R_1$ for  $n=1,..., N$, and $r_{m, j^\prime}=R_0$ or $R_1$ for $m=1,..., N$. Therefore, row $i$ and column $j^\prime$ are a non-SP row and column, respectively.

By contrast, if $x_{i,j^\prime}=1$, according to (\ref{eq:readout-eg2}), the readout resistance $r_{i, n}, n\neq j, j^\prime$ can be one of three situations: $r_{i, n}=R_1$ if $x_{i, n}=1$, or $r_{i, n}=R_0^\prime$ if $x_{i, n}=0, x_{i^\prime, n}=1$, or $r_{i, n}=R_0$ if $x_{i, n}=0, x_{i^\prime, n}=0$. Since in the third situation, cell $(i, n)$ will not be a critical cell, we conclude that row $i$ is either a non-SP row (only the first and third situations occur) or a complete-SP row (the second situation exists). Since the former happens with probability $(q+(1-q)^2)^{N-2}$, the later happens with probability $1-(q+(1-q)^2)^{N-2}$. A similar proof applies to SF column $j^\prime$.\myQED

Based on Lemma~\ref{lem:7}, applying Bayes' rule we obtain the following lemma.
 \begin{lemma}\label{lem:8}
 For Double-Selector Failure with SFs at cells $(i, j)$ and $(i^\prime, j^\prime)$, if SF row $i$ or SF column $j^\prime$ is a complete-SP row or column, $x_{i,j^\prime}=1$ holds.
 If SF row $i$ and SF column $j^\prime$ are a non-SP row and column, $x_{i,j^\prime}=0$ holds with probability $1-q(q+(1-q)^2)^{2N-4}/(1-q)$. Similar conclusions hold for SF row $i^\prime$ and SF column $j$ and $x_{i^\prime,j}$.
\end{lemma}

From Lemmas~\ref{lem:7} and \ref{lem:8}, we can demonstrate a one-to-one correspondence between intersectional entries $x_{i,j^\prime}, x_{i^\prime,j}$, and the asymptotical SP row or column types of the SF rows or columns as $N\rightarrow\infty$ in Fig.~\ref{fig:SFrowSP}.

\begin{figure}[t]
\includegraphics[width=
3.5 in]{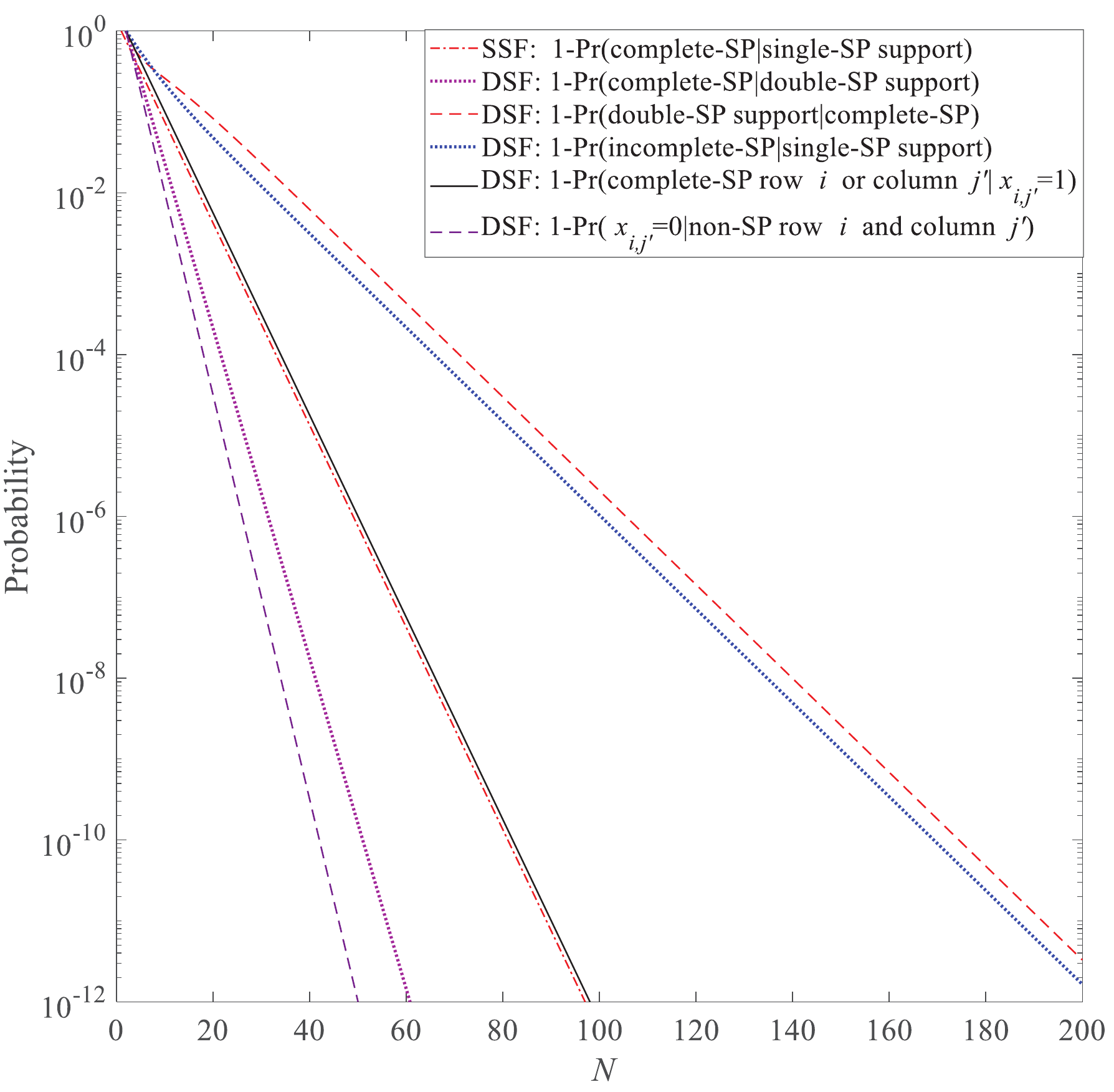}
\centering
\caption{Illustration of conditional probabilities when $q=0.5$ as array size $N$ increases. SSF: Single-Selector Failure. DSF: Double-Selector Failure. Pr$(E_2|E_1)$: the probability of event $E_2$ under the condition of event $E_1$. Pr(complete-SP row $i$ or column $j^\prime|x_{i,j^\prime}=1$): the probability of the SF row $i$ or SF column $j^\prime$ being a complete row or column when intersectional entry $x_{i,j^\prime}=1$. Pr($x_{i,j^\prime}=0|$non-SP row $i$ and column $j^\prime$): the probability of SF row $i$ (or SF column $j^\prime$) being a complete row or column when intersectional entry $x_{i,j^\prime}=1$.} \label{fig:Probability-SP}
\end{figure}

We illustrate these conditional probabilities studied in Lemmas~\ref{lem:4}-\ref{lem:8} when $q=0.5$ in Fig.~\ref{fig:Probability-SP}. It shows that when $N>100$, all the values of (1-probability) are less than $10^{-6}$, which implies the one-to-one corresponding relationships shown in TABLE~\ref{tab:SP-type} and Fig.~\ref{fig:SFrowSP}.

\section{SF-Row and -Column $\fat{X}_{sf}$ Detection}\label{sec:SF-detection}
In this section, we demonstrate detection of SF rows and columns $\fat{X}_{sf}$ and SF locations in an array. We assume array size $N$ is large enough so that the one-to-one correspondences between SP-row and -column types and the number of SP supports hold.
Our idea is that we first detect the SP type of each row or column, based on which we can detect the SF pattern, SF locations, and SF rows and columns.

To simplify our notation, we always use $(i, j)$ and $(i, j), (i^\prime, j^\prime)$ to denote the SF locations for Single-Selector Failure and Double-Selector Failure, respectively.

\subsection{SP-Row and -Column Type Detection and SF Pattern Determination}
Let $\fat{x}^-_m$ and $\fat{x}^\shortmid_m$ denote the $m$-th row and column of array $\fat{X}$, respectively. Similarly, let $\fat{y}^-_m$ and $\fat{y}^\shortmid_m$ denote the $m$-th row and column of array $\fat{Y}$. Let $\fat{s}^\shortmid=(s^\shortmid_1,...,s^\shortmid_N)$ and $\fat{s}^-=(s^-_1,...,s^-_N), s^\shortmid_n, s_m^-\in\{0, \frac{1}{2}, 1\}, m, n=1,...,N$, be SP-column and SP-row type-indicator vectors of the memory array, respectively, with
\begin{eqnarray}
s^\shortmid_n=\begin{cases} 0&\mbox{if column $n$ is a non-SP column}\\  \frac{1}{2}&\mbox{if column $n$ is an incomplete-SP column}\\ 1 &\mbox{if column $n$ is a complete-SP column} \end{cases}
\end{eqnarray}
\begin{eqnarray}
s_m^-=\begin{cases} 0&\mbox{if row $m$ is a non-SP row}\\  \frac{1}{2}&\mbox{if row $m$ is an incomplete-SP row}\\ 1 &\mbox{if row $m$ is a complete-SP row}. \end{cases}
\end{eqnarray}
In the following, we detect $\fat{s}^\shortmid, \fat{s}^-$ based on readout signal array $\fat{Y}$.

Since the probability density function (PDF) of readout signal $y_{m,n}$ in (\ref{eq:readout}) is a mixture of three Gaussian PDFs with means of $R_1, R_0$, and $R_0^\prime$ and the same variance $\sigma^2$, in the following, we use
\begin{eqnarray}
\rho_{y_{m,n}}(a,b,c)=ae^{-\frac{(y_{m,n}-R_1)^2}{2\sigma^2}}+be^{-\frac{(y_{m,n}-R_0)^2}{2\sigma^2}}+ce^{-\frac{(y_{m,n}-R_0^\prime)^2}{2\sigma^2}}
\end{eqnarray}
to denote this function, where $0\leqslant a,b,c\leqslant 1$ with $a+b+c=1$ are the priori probabilities of each component.

 Since the detector does not know the SF pattern, it presupposes Double-Selector Failure at the beginning of detection. Since $\fat{s}^\shortmid$ and $\fat{s}^-$ are ternary vectors, we need a two-step log-likelihood ratio (LLR) calculation to determine them.

 \emph{[Step 1:]} The SP type of column $n$, i.e., $s_n^\shortmid$, is detected based on $\fat{y}^\shortmid_n$ for $n=1,...,N$. We first calculate the following LLR:
\begin{eqnarray}
L_1(s_n^\shortmid)\!\!\!\!\!\!\!&&=\log\frac{\textrm{Pr}(\fat{y}^\shortmid_n|s_n^\shortmid=1/2)}{\textrm{Pr}(\fat{y}^\shortmid_n|s_n^\shortmid=0)}\\
\!\!\!\!\!\!\!&&=\log\frac{\prod_{m=1}^N\textrm{Pr}(y_{m,n}|s_n^\shortmid=1/2)}{\prod_{m=1}^N\textrm{Pr}(y_{m,n}|s_n^\shortmid=0)}\\
\!\!\!\!\!\!\!&&=\sum_{m=1}^N\log\frac{\rho_{y_{m,n}}(q,(1-q)^2,(1-q)q)}{\rho_{y_{m,n}}(q,1-q,0)}\label{eq:SP-type1}
\end{eqnarray}
where (\ref{eq:SP-type1}) is because if column $n$ is an incomplete-SP column (containing one SP support),
with probability $q,(1-q)^2,(1-q)q$, cell $(m, n)$ is an LRS, HRS, SP cell, and if it is a non-SP column (not containing SP support),
with probability $q,1-q$,  cell $(m, n)$ is an LRS, HRS cell.

Similarly, we have the LLR of $s_m^-, m=1,...,N$:
\begin{eqnarray}
L_1(s_m^-)\!\!\!\!\!\!\!&&=\log\frac{\textrm{Pr}(\fat{y}^-_m|s_m^-=1/2)}{\textrm{Pr}(\fat{y}^-_n|s_m^-=0)}\\
\!\!\!\!\!\!\!&&=\sum_{n=1}^N\log\frac{\rho_{y_{m,n}}(q,(1-q)^2,(1-q)q)}{\rho_{y_{m,n}}(q,1-q,0)}.\nonumber
\end{eqnarray}

We do the first step hard decision as follows:
\begin{eqnarray}
\tilde{s}_n^\shortmid&&\!\!\!\!\!\!\!=\begin{cases} 0&\mbox{if $L_1(s_n^\shortmid)<0$}\\  1/2&\mbox{if $L_1(s_n^\shortmid)\geq0$}\end{cases}
\end{eqnarray}
\begin{eqnarray}
\tilde{s}_m^-&&\!\!\!\!\!\!\!=\begin{cases} 0&\mbox{if $L_1(s_m^-)<0$}\\  1/2&\mbox{if $L_1(s_m^-)\geq0$}. \end{cases}
\end{eqnarray}

 Using $\tilde{s}_n^\shortmid$ and $\tilde{s}_m^-$ as a priori information, we do the second step LLR calculation.

 \emph{[Step 2:]} For each $n$ with $\tilde{s}_n^\shortmid=1/2, 1\leq n\leq N$, we have
 \begin{eqnarray}
L_2(s_n^\shortmid)\!\!\!\!\!\!\!&&=\log\frac{\textrm{Pr}(\fat{y}^\shortmid_n|s_n^\shortmid=1, \fat{s}^-= \tilde{\fat{s}}^-)}{\textrm{Pr}(\fat{y}^\shortmid_n|s_n^\shortmid=1/2, \fat{s}^-=\tilde{\fat{s}}^-)}\\
\!\!\!\!\!\!\!&&=\log\frac{\prod_{m=1}^N\textrm{Pr}(y_{m,n}|s_n^\shortmid=1, s_m^-=\tilde{s}_m^-)}{\prod_{m=1}^N\textrm{Pr}(y_{m,n}|s_n^\shortmid=1/2, s_m^-=\tilde{s}_m^-)}\\
\!\!\!\!\!\!\!&&=\sum_{m=1}^N2\tilde{s}_m^-\log\frac{\rho_{y_{m,n}}(q,0,1-q)}{\rho_{y_{m,n}}(q,(1-q)/2,(1-q)/2)}\label{eq:SP-type2}
\end{eqnarray}
where (\ref{eq:SP-type2}) is because if $\tilde{s}_m^-=0$, $\textrm{Pr}(y_{m,n}|s_n^\shortmid=1, s_m^-=\tilde{s}_m^-)$=$\textrm{Pr}(y_{m,n}|s_n^\shortmid=1/2, s_m^-=\tilde{s}_m^-)$=$\rho_{y_{m,n}}(q,1-q,0)$ and if $\tilde{s}_m^-=1/2$,
$\textrm{Pr}(y_{m,n}|s_n^\shortmid=1, s_m^-=\tilde{s}_m^-)$=$\rho_{y_{m,n}}(q,0,1-q)$ and $\textrm{Pr}(y_{m,n}|s_n^\shortmid=1/2, s_m^-=\tilde{s}_m^-)$=$\rho_{y_{m,n}}(q,(1-q)/2,(1-q)/2)$.
Similarly, we have the LLR of $s_m^-, m=1,...,N$,
\begin{eqnarray}
L_2(s_m^-)\!\!\!\!\!\!\!&&=\log\frac{\textrm{Pr}(\fat{y}^-_m|s_m^-=1, \tilde{\fat{s}}^-)}{\textrm{Pr}(\fat{y}^-_n|s_m^-=1/2, \tilde{\fat{s}}^-)}\\
\!\!\!\!\!\!\!&&=\sum_{n=1}^N2\tilde{s}_n^-\log\frac{\rho_{y_{m,n}}(q,0,1-q)}{\rho_{y_{m,n}}(q,(1-q)/2,(1-q)/2)}.\nonumber
\end{eqnarray}

Based on $L_1$ and $L_2$, we have the final hard decision for $s_n^\shortmid, s_m^-, m,n=1,...,N$,
\begin{eqnarray}
\hat{s}_n^\shortmid&&\!\!\!\!\!\!\!=\begin{cases} 0&\mbox{if $L_1(s_n^\shortmid)<0$}\\  1/2&\mbox{if $L_1(s_n^\shortmid)\geq0, L_2(s_n^\shortmid)<0$}\\  1&\mbox{if $L_1(s_n^\shortmid)\geq0, L_2(s_n^\shortmid)\geq0$}\end{cases}
\label{eq:hard1}
\end{eqnarray}
\begin{eqnarray}
\hat{s}_m^-&&\!\!\!\!\!\!\!=\begin{cases} 0&\mbox{if $L_1(s_m^-)<0$}\\  1/2&\mbox{if $L_1(s_m^-)\geq0, L_2(s_m^-)<0$} \\  1&\mbox{if $L_1(s_m^-)\geq0, L_2(s_m^-)\geq0$}. \end{cases}
\label{eq:hard2}
\end{eqnarray}

The SF pattern can be determined based on the estimated SP-row- or -column-type vector. If $\hat{s}_n^\shortmid=0$ holds for $n=1,...,N$, no SF is declared, or otherwise, if $\hat{s}_n^\shortmid\in\{0, 1\}$ holds for $n=1,...,N$, Single-Selector Failure is declared, otherwise, if $\hat{s}_n^\shortmid\in\{0, 1, 2\}$ holds for $n=1,...,N$, Double-Selector Failure is declared.

Once the array is detected with a non-zero number of SFs, we need to further detect the SF locations and the SF rows and columns.
We propose detection methods for both Single-Selector Failure and Double-Selector Failure.
\subsection{SF Row and Column Detection: Single-Selector Failure}

According to Lemma~\ref{lem:SFSPtype}, the SF row or column can only be a
non-SP row or column, thus, we only need to find the SF-row and -column indices from index sets $\{m|\hat{s}^-_m=0, 1\leqslant m\leqslant N\}$ and $\{n|\hat{s}^\shortmid_n=0, 1\leqslant n\leqslant N\}$, respectively.
Since for Single-Selector Failure, the SP-row or -column type-indicator vector, which describes the SP-support number in each row or column,
is exactly the SF-column or -row vector (the only mismatch is at the SF center, which should be logical-1),
 we can determine the SF-row and -column indices by finding the row and column of $\fat{Y}$, which have maximum correlations with $\hat{\fat{s}}^\shortmid$ and $\hat{\fat{s}}^-$:
\begin{eqnarray}
\hat{i}=\!\!\!\!\!\!&&\arg\max_{m\in\{m|\hat{s}^-_m=0, 1\leqslant m\leqslant N\}}\log \textrm{Pr}\left(\fat{y}^-_m|\fat{x}^-_m=\hat{\fat{s}}^\shortmid\right)\\
=\!\!\!\!\!\!&&\arg\max_{m\in\{m|\hat{s}^-_m=0, 1\leqslant m\leqslant N\}}\sum_{n=1}^N\log \exp\left(-\frac{(y_{m,n}-R_{\hat{s}^\shortmid_n})^2}{2\sigma^2}\right)\\
=\!\!\!\!\!\!&&\arg\min_{m\in\{m|\hat{s}^-_m=0, 1\leqslant m\leqslant N\}}\sum_{n=1}^N(y_{m,n}-R_{\hat{s}^\shortmid_n})^2\\
\hat{j}=\!\!\!\!\!\!&&\arg\min_{n\in\{n|\hat{s}^\shortmid_n=0, 1\leqslant n\leqslant N\}}\sum_{m=1}^N(y_{m,n}-R_{\hat{s}^-_m})^2.
\end{eqnarray}
Simultaneously, the SF row and column can be determined as
\begin{eqnarray}
\hat{x}_{\hat{i},n}&&\!\!\!\!\!\!\!=\begin{cases} \hat{s}_n^\shortmid&\mbox{if $n\neq \hat{j}$}\\  1&\mbox{if $n=\hat{j}$}\end{cases}
\label{eq:SFrow1}
\end{eqnarray}
\begin{eqnarray}
\hat{x}_{m,\hat{j}}&&\!\!\!\!\!\!\!=\begin{cases} \hat{s}_m^-&\mbox{if $m\neq \hat{i}$}\\  1&\mbox{if $m=\hat{i}$} \end{cases}
\label{eq:SFcolumn1}
\end{eqnarray}
for $m, n=1,...,N$.

\subsection{SF-Row and -Column Detection: Double-Selector Failure}
We first determine the locations of the SFs based on estimated SP-row and -column type-indicator vectors $\hat{\fat{s}}^-$ and $\hat{\fat{s}}^\shortmid$, and then we detect the SF rows and columns.

Since for Double-Selector Failure the SP-row or -column type-indicator vectors specify the number of SP supports in the rows or columns, we have equations $\fat{s}^\shortmid\approx\frac{\fat{x}^-_i+\fat{x}^-_{i^\prime}}{2}$ and $\fat{s}^-\approx\frac{\fat{x}^\shortmid_j+\fat{x}^\shortmid_{j^\prime}}{2}$ (the only differences are at the intersections of the SF rows and columns, which will be separately detected). Therefore,
the SF-row indices can be detected based on ML criteria $(\hat{i}, \hat{i}^\prime)=\arg\max_{(m, m^\prime)}\textrm{Pr}(\fat{y}_m^-, \fat{y}_{m^\prime}^-|\fat{x}^-_m+\fat{x}^-_{m^\prime}=2\hat{\fat{s}}^\shortmid)$. Since the joint ML detection for $i, i^\prime$ leads to high complexity,
we use a simplified detection scheme based on a separate ML criterion. Specifically, we find SF-row index $\hat{i}=\arg\max_m\textrm{Pr}(\fat{y}_m^-|\fat{x}^-_m+\fat{x}^-_{m^\prime}=2\hat{\fat{s}}^\shortmid)$ by regarding $\fat{x}^-_{m^\prime}$ as a random variable, and find $\hat{i}^\prime=\arg\max_{m^\prime}\textrm{Pr}(\fat{y}_{m^\prime}^-|\fat{x}^-_m+\fat{x}^-_{m^\prime}=2\hat{\fat{s}}^\shortmid)$ by regarding $\fat{x}^-_m$ as a random variable. We can solve these two problems together. Let $\textrm{Max}2$ be a function of finding the largest two values of a function.
Since according to Lemma~\ref{lem:7}, the SF rows and columns can either be non-SP rows and columns or complete-SP rows and columns, we find an SF-row index candidate set
\begin{eqnarray}
\{i_1, i_2\}\!\!\!\!\!\!\!\!\!&&=\arg\mathop{\textrm{Max}2}_{m\in\{m|\hat{s}^-_m\neq1/2, 1\leqslant m\leqslant N\}}\log \textrm{Pr}\left(\fat{y}^-_m|\fat{x}^-_m+\fat{x}^-_{m^\prime}=2\hat{\fat{s}}^\shortmid\right)\\
\!\!\!\!\!\!&&=\arg\mathop{\textrm{Max}2}_{m\in\{m|\hat{s}^-_m\neq1/2, 1\leqslant m\leqslant N\}}\sum_{n=1}^N\log \textrm{Pr}\left(y_{m,n}|x_{m,n}+x_{m^\prime, n}=2\hat{s}_n^\shortmid\right)\nonumber\\
\!\!\!\!\!\!&&=\arg\mathop{\textrm{Max}2}_{m\in\{m|\hat{s}^-_m\neq1/2, 1\leqslant m\leqslant N\}}\sum_{n=1}^N\log\nonumber\\
\!\!\!\!\!\!&&\ \ \ \ \times \rho_{y_{m,n}}(\hat{s}^\shortmid_n,1-\hat{s}^\shortmid_n-\frac{1}{2}\delta_{\hat{s}^\shortmid_n,\frac{1}{2}}\delta_{\hat{s}^-_m,1},\frac{1}{2}\delta_{\hat{s}^\shortmid_n,\frac{1}{2}}\delta_{\hat{s}^-_m,1})\label{eq:i12}
\end{eqnarray}
where $\delta_{a,b}$ is the Kronecker delta function with $\delta_{a,b}=1$ if $a=b$ and  $\delta_{a,b}=0$ if $a\neq b$.
Equation (\ref{eq:i12}) is because when $\hat{s}_n^\shortmid\in\{0, 1\}$, we have $x_{m,n}=x_{m^\prime, n}=\hat{s}_n^\shortmid$ leading to $\textrm{Pr}(y_{m,n})=\rho_{y_{m,n}}(\hat{s}^\shortmid_n,1-\hat{s}^\shortmid_n, 0)$, and when $\hat{s}_n^\shortmid=1/2$, $\textrm{Pr}(x_{m,n}=0, x_{m^\prime, n}=1)=\textrm{Pr}(x_{m,n}=1, x_{m^\prime, n}=0)=1/2$ leading to $\textrm{Pr}(y_{m,n})=\rho_{y_{m,n}}(\frac{1}{2},\frac{1}{2}-\frac{1}{2}\delta_{\hat{s}^-_m,1},\frac{1}{2}\delta_{\hat{s}^-_m,1})$.
Similarly, we can find an SF-column index candidate set
\begin{eqnarray}
\{j_1, j_2\}\!\!\!\!\!\!\!\!\!&&=\arg\mathop{\textrm{Max}2}_{n\in\{n|\hat{s}^\shortmid_n\neq1/2, 1\leqslant n\leqslant N\}}\sum_{m=1}^N\log\nonumber\\
\!\!\!\!\!\!&&\ \  \times \rho_{y_{m,n}}(\hat{s}^-_m,1-\hat{s}^-_m-\frac{1}{2}\delta_{\hat{s}^-_m,\frac{1}{2}}\delta_{\hat{s}^\shortmid_n,1},\frac{1}{2}\delta_{\hat{s}^-_m,\frac{1}{2}}\delta_{\hat{s}^\shortmid_n,1}).
\end{eqnarray}

Since there are two SFs at the intersections of SF rows and SF columns, we need to further determine whether they are at $(i_1, j_1), (i_2, j_2)$, referred to as hypothesis $H_0$, or at $(i_1, j_2), (i_2, j_1)$, referred to as hypothesis $H_1$. In other words, if we assume $\hat{i}=i_1, \hat{i}^\prime=i_2$, we need to further determine whether $\hat{j}=j_1, \hat{j}^\prime=j_2$ or $\hat{j}=j_2, \hat{j}^\prime=j_1$ holds. For given $m, n, 0<m, n\leq N$, if $\hat{s}^-_m\in\{0, 1\}$, we can immediately recover the corresponding entries of SF rows as $\hat{x}_{i_1,n}=\hat{x}_{i_2,n}=\hat{s}^\shortmid_n$. However, if $\hat{s}^-_n=1/2$, the corresponding entries in the SF rows are uncertain, i.e., $(\hat{x}_{i_1,n},\hat{x}_{i_2,n})=(0, 1)$ or $(1, 0)$. Since we already know the SF-row indices, we can estimate these uncertain entries based on their readout signal. Specifically,
for $1\leqslant n\leqslant N$ with $\hat{s}^\shortmid_n=1/2$, we calculate an LLR:
\begin{eqnarray}
L_1(x_{i_1,n}, x_{i_2,n})&&\!\!\!\!\!\!\!\!\!\!\!\!=\log\frac{\textrm{Pr}(y_{i_1,n},y_{i_2,n}|x_{i_1,n}\!=\!0,x_{i_2,n}\!=\!1,s^-_{i_1}\!=\!\hat{s}^-_{i_1},s^-_{i_2}\!=\!\hat{s}^-_{i_2})}
{\textrm{Pr}(y_{i_1,n},y_{i_2,n}|x_{i_1,n}\!=\!1,x_{i_2,n}\!=\!0,s^-_{i_1}\!=\!\hat{s}^-_{i_1},s^-_{i_2}\!=\!\hat{s}^-_{i_2})}\nonumber\\
&&\!\!\!\!\!\!\!\!\!\!\!\!=\log\frac{\exp\left(-\frac{\left(y_{i_1,n}-R_0({\hat{s}^-_{i_1}})\right)^2+\left(y_{i_2,n}-R_1\right)^2}{2\sigma^2}\right)}{\exp\left(-\frac{\left(y_{i_1,n}-R_1\right)^2+\left(y_{i_2,n}-R_0({\hat{s}^-_{i_2}})\right)^2}{2\sigma^2}\right)}\nonumber\\
&&\!\!\!\!\!\!\!\!\!\!\!\!=\left(2y_{i_1,n}\left(R_0({\hat{s}^-_{i_1}})-R_1\right)-2y_{i_2,n}\left(R_0({\hat{s}^-_{i_2}})-R_1\right)\right.\nonumber\\
&&\!\!\!\!\!\!\!\!\!\!\!\! \ \ \ \ \ \left.+R_0({\hat{s}^-_{i_2}})^2-R_0({\hat{s}^-_{i_1}})^2\right)/(2\sigma^2)\label{eq:L1row}
\end{eqnarray}
where we used $\hat{s}^-_{i_1}, \hat{s}^-_{i_2}\in\{0, 1\}$ in the above derivation.
Similarly, for each $1\leqslant m\leqslant N$ with $\hat{s}_m^-=1/2$, we calculate the LLR for uncertain bits in SF columns as
\begin{eqnarray}
L_1(x_{m,j_1}, x_{m,j_2})&&\!\!\!\!\!\!\!\!\!\!\!\!=\log\frac{\textrm{Pr}(y_{m,j_1},y_{m,j_2}|x_{m,j_1}\!=\!0,x_{m,j_2}\!=\!1,\hat{s}^\shortmid_{j_1},\hat{s}^\shortmid_{j_2})}
{\textrm{Pr}(y_{m,j_1},y_{m,j_2}|x_{m,j_1}\!=\!1,x_{m,j_2}\!=\!0,\hat{s}^\shortmid_{j_1},\hat{s}^\shortmid_{j_2})}\nonumber\\
&&\!\!\!\!\!\!\!\!\!\!\!\!=\left(2y_{m,j_1}\left(R_0({\hat{s}^\shortmid_{j_1}})-R_1\right)-2y_{m,j_2}\left(R_0({\hat{s}^\shortmid_{j_2}})-R_1\right)\right.\nonumber\\
&&\!\!\!\!\!\!\!\!\!\!\!\! \ \ \ \ \ \left.+R_0({\hat{s}^\shortmid_{j_2}})^2-R_0({\hat{s}^\shortmid_{j_1}})^2\right)/(2\sigma^2).\label{eq:L1column}
\end{eqnarray}

Based on $\hat{\fat{s}}^-, \hat{\fat{s}}^\shortmid$ and $L_1(x_{i_1,n}, x_{i_2,n}), L_1(x_{m,j_1}, x_{m,i_2})$, we do the first step hard decision for the SF rows and columns
\begin{equation}\label{eq:est1}
(\tilde{x}_{i_1,n},\tilde{x}_{i_2,n})=\begin{cases} (0, 0)&\mbox{if $\hat{s}_n^\shortmid=0$}\\ (1, 1)&\mbox{if $\hat{s}_n^\shortmid=1$}\\
(0, 1)&\mbox{if $\hat{s}_n^\shortmid=1/2, L_1(x_{i_1,n}, x_{i_2,n})>0$}\\
(1, 0)&\mbox{if $\hat{s}_n^\shortmid=1/2, L_1(x_{i_1,n}, x_{i_2,n})\leq0$} \end{cases}
\end{equation}
for $n =1,...,N$, and
\begin{equation}\label{eq:est2}
(\tilde{x}_{m,j_1},\tilde{x}_{m,j_2})=\begin{cases} (0, 0)&\mbox{if $\hat{s}_m^-=0$}\\ (1, 1)&\mbox{if $\hat{s}_m^-=1$}\\
(0, 1)&\mbox{if $\hat{s}_m^-=1/2, L_1(x_{m,j_1}, x_{m,i_2})>0$}\\
(1, 0)&\mbox{if $\hat{s}_m^-=1/2, L_1(x_{m,j_1}, x_{m,i_2})\leq0$} \end{cases}
\end{equation}
for $m =1,...,N$.

\subsubsection{Cases (1), (2), and (3) in Fig.~\ref{fig:SFrowSP}}

When the SP types of the SF rows and columns are identical to one of cases of (1), (2), or (3) in Fig.~\ref{fig:SFrowSP}, i.e., at least one of the SF rows or columns is a non-SP row or column,
the SF locations are easy to determine.

 For case (1) in Fig.~\ref{fig:SFrowSP}, since the two SF cells are with LRS while the
other two are with HRS, we can calculate the following LLR based on the readback signals of the intersectional cells:
\begin{eqnarray}
L_H&&\!\!\!\!\!\!\!\!\!\!\!\!=\log\frac{\textrm{Pr}(y_{i_1,j_1}, y_{i_2,j_2},y_{i_1,j_2},y_{i_2,j_1}|H_0)}{\textrm{Pr}(y_{i_1,j_1}, y_{i_2,j_2},y_{i_1,j_2},y_{i_2,j_1}|H_1)}\nonumber\\
&&\!\!\!\!\!\!\!\!\!\!\!\!=\log\frac{\exp\left(-\frac{(y_{i_1,j_1}-R_1)^2+(y_{i_2,j_2}-R_1)^2+(y_{i_1,j_2}-R_0)^2+(y_{i_2,j_1}-R_0)^2}{2\sigma^2}\right)}{\exp\left(-\frac{(y_{i_1,j_1}-R_0)^2+(y_{i_2,j_2}-R_0)^2+(y_{i_1,j_2}-R_1)^2+(y_{i_2,j_1}-R_1)^2}{2\sigma^2}\right)}\nonumber\\
&&\!\!\!\!\!\!\!\!\!\!\!\!=(y_{i_1,j_1}+ y_{i_2,j_2}-y_{i_1,j_2}-y_{i_2,j_1})(R_1-R_0)/\sigma^2.
\end{eqnarray}
If $L_H>0$, we say that $H_0$ is correct, otherwise, we say $H_1$ is correct.

For cases (2) and (3) in Fig.~\ref{fig:SFrowSP}, the SF locations can be determined by observing the SP types of the SF rows or columns. They are located at the intersectional cells of the SF rows and columns with different SP types.

For cases (1), (2), and (3) in Fig.~\ref{fig:SFrowSP}, since at least one of the SF rows or columns is a non-SP row or column, the first step hard decisions (\ref{eq:est1}) and (\ref{eq:est2}) for the SF rows or columns are very accurate. Thus, we can use them as our final hard decision  $(\hat{x}_{i_1,n},\hat{x}_{i_2,n})=(\tilde{x}_{i_1,n},\tilde{x}_{i_2,n})$ and  $(\hat{x}_{m,j_1},\hat{x}_{m,j_2})=(\tilde{x}_{m,j_1},\tilde{x}_{m,j_2}), m, n=1,...,N$.

\subsubsection{Case (4) in Fig.~\ref{fig:SFrowSP}}
For case (4) in Fig.~\ref{fig:SFrowSP}, since all the SF rows and columns are complete-SP rows and columns, we cannot determine the SF locations in similar ways to those for cases (1), (2), and (3), and the hard decisions given by (\ref{eq:est1}) and (\ref{eq:est2}) for the uncertain entries in the SF rows and columns are not accurate. We need to propose a new method to determine the SF locations and refine the detection for the uncertain entries.

We propose a simple method to determine the SF locations based on the hard decision values for the resistances.
We assume that each non-SF entry has a hard decision value $\tilde{r}_{m,n}$ based on the minimum Euclidean distance rule:
\begin{equation}
\tilde{r}_{m,n}=\arg\min_{r\in\{R_1, R_0, R_0^\prime\}}|y_{m,n}-r|
\end{equation}
for $0\leq m, n\leq N$ and $m\neq i_1, i_2, n\neq j_1, j_2$.
  Since a critical cell at the intersection of an SP-supported row and an SP-supported column that is caused by the same SF
is either an LRS cell or an SP cell (without an HRS cell), we call it a contradiction if such an intersectional cell has HRS $(R_0)$. We can check the two hypotheses $H_0$ and $H_1$ using the hard decision values for the resistances and find the one with fewer contradictions.
 Specifically, we calculate the difference between the numbers of contradictions created by these two hypotheses:
 \begin{eqnarray}
 L_H&&\!\!\!\!\!\!\!\!\!\!=\sum_{m\neq i_1, i_2}\sum_{n\neq j_1, j_2}\left(\tilde{x}_{i_1,n}\tilde{x}_{m,j_2}+\tilde{x}_{i_2,n}\tilde{x}_{m,j_1}\right)\delta_{\tilde{r}_{m,n},R_0}\nonumber\\
 &&\!\!\!\!\!\!\!\!\!\!\ \ \ \ \ \ \ \ \ -\sum_{m\neq i_1, i_2}\sum_{n\neq j_1, j_2}\left(\tilde{x}_{i_1,n}\tilde{x}_{m,j_1}+\tilde{x}_{i_2,n}\tilde{x}_{m,j_2}\right)\delta_{\tilde{r}_{m,n},R_0}\nonumber\\
 &&\!\!\!\!\!\!\!\!\!\!=\sum_{m\neq i_1, i_2}\sum_{n\neq j_1, j_2}(\tilde{x}_{i_1,n}-\tilde{x}_{i_2,n})(\tilde{x}_{m,j_2}-\tilde{x}_{m,j_1})\delta_{\tilde{r}_{m,n},R_0}.
 \end{eqnarray}
If $L_H>0$, we say that $H_0$ is correct, otherwise, we say that $H_1$ is correct.

\begin{figure}[t]
\includegraphics[width=
2.8 in]{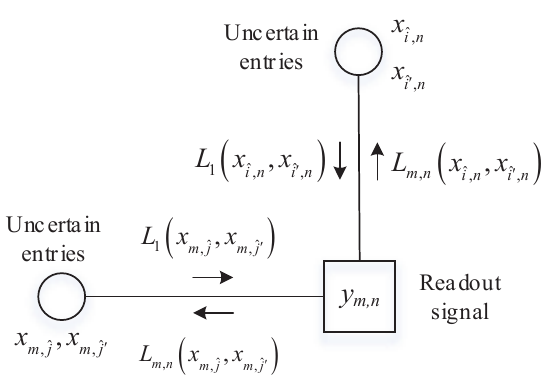}
\centering
\caption{Illustration of message calculation for uncertain entries} \label{fig:uncertain}
\end{figure}

 After knowing the SF locations, we can provide a more reliable estimation for the SF rows and columns. Suppose $(x_{\hat{i},n},x_{\hat{i}^\prime,n})\in\{(0,1), (1,0)\}$ and $(x_{m,\hat{j}},x_{m,\hat{j}^\prime})\in\{(0,1),(1,0)\}$ are two uncertain pairs. Since the readout signal $y_{m,n}$ is related to both $(x_{\hat{i},n},x_{\hat{i}^\prime,n})$ and $(x_{m,\hat{j}},x_{m,\hat{j}^\prime})$, we can use the LLRs derived in (\ref{eq:L1row}) and (\ref{eq:L1column}) as a priori and refine the estimation based on $y_{m,n}$.
We first calculate the following probability:
\begin{eqnarray}
\textrm{Pr}(y_{m,n}|x_{\hat{i},n},x_{\hat{i}^\prime,n})&&\!\!\!\!\!\!\!\!\!\!\!\!=\sum_{x_{m,\hat{j}},x_{m,\hat{j}^\prime}}\sum_{x_{m,n}}\textrm{Pr}(x_{m,\hat{j}},x_{m,\hat{j}^\prime})\textrm{Pr}(x_{m,n})\nonumber\\
&&\!\!\!\!\!\!\!\!\!\!\!\!\ \ \ \times\textrm{Pr}\left(y_{m,n}|x_{m,n},x_{\hat{i},n}x_{m,\hat{j}}, x_{\hat{i}^\prime,n}x_{m,\hat{j}^\prime}\right)\nonumber\\
&&\!\!\!\!\!\!\!\!\!\!\!\!=\sum_{x_{m,\hat{j}},x_{m,\hat{j}^\prime}}\textrm{Pr}(x_{m,\hat{j}},x_{m,\hat{j}^\prime})\sum_{x_{m,n}}\textrm{Pr}(x_{m,n})\nonumber\\
&&\!\!\!\!\!\!\!\!\!\!\!\!\ \ \ \times\textrm{Pr}\left(y_{m,n}|r_{m,n}=R_{x_{m,n}}\left(x_{\hat{i},n}x_{m,\hat{j}}\bigcup x_{\hat{i}^\prime,n}x_{m,\hat{j}^\prime}\right)\right)\nonumber
\end{eqnarray}
which leads to
\begin{eqnarray}
\textrm{Pr}(y_{m,n}|x_{\hat{i},n}=0,x_{\hat{i}^\prime,n}=1)&&\!\!\!\!\!\!\!\!\!\!\!\!=\textrm{Pr}(x_{m,\hat{j}}=0,x_{m,\hat{j}^\prime}=1)\rho_{y_{m,n}}(q,0,1-q)\nonumber\\
&&\!\!\!\!\!\!\!\!\!\!\!\! + \textrm{Pr}(x_{m,\hat{j}}=1,x_{m,\hat{j}^\prime}=0)\rho_{y_{m,n}}(q,1-q,0)\nonumber
\end{eqnarray}
and
\begin{eqnarray}
\textrm{Pr}(y_{m,n}|x_{\hat{i},n}=1,x_{\hat{i}^\prime,n}=0)&&\!\!\!\!\!\!\!\!\!\!\!\!=\textrm{Pr}(x_{m,\hat{j}}=0,x_{m,\hat{j}^\prime}=1)\rho_{y_{m,n}}(q,1-q,0)\nonumber\\
&&\!\!\!\!\!\!\!\!\!\!\!\!  + \textrm{Pr}(x_{m,\hat{j}}=1,x_{m,\hat{j}^\prime}=0)\rho_{y_{m,n}}(q,0,1-q)\nonumber.
\end{eqnarray}
Then we have the following LLR
\begin{eqnarray}
&&\!\!\!\!\!\!\!\!\!\!\!\!L_{m,n}(x_{\hat{i},n},x_{\hat{i}^\prime,n})=\log\frac{\textrm{Pr}(y_{m,n}|x_{\hat{i},n}=0,x_{\hat{i}^\prime,n}=1)}
{\textrm{Pr}(y_{m,n}|x_{\hat{i},n}=1,x_{\hat{i}^\prime,n}=0)}\nonumber\\
&&\!\!\!\!\!\!\!\!\!\!\!\!=\log\frac{\exp(L_1(x_{m,\hat{j}},x_{m,\hat{j}^\prime}))\rho_{y_{m,n}}(q,0,1-q)\!+\!\rho_{y_{m,n}}(q,1-q,0)}{\exp(L_1(x_{m,\hat{j}},x_{m,\hat{j}^\prime}))\rho_{y_{m,n}}(q,1-q,0)\!+\!\rho_{y_{m,n}}(q,0,1-q)}\nonumber
\end{eqnarray}
where we replaced $\log\frac{\textrm{Pr}(x_{m,\hat{j}}=0,x_{m,\hat{j}^\prime}=1)}{\textrm{Pr}(x_{m,\hat{j}}=1,x_{m,\hat{j}^\prime}=0)}$ by $L_1(x_{m,\hat{j}},x_{m,\hat{j}^\prime})$ in the above derivation. Figure~\ref{fig:uncertain} illustrates the message calculation for uncertain entries.

Based on the readout signal of $n$-th column $\fat{y}_n^\shortmid$, we have
\begin{eqnarray}
&&\!\!\!\!\!\!\!\!\!\!\!\!L_2(x_{\hat{i},n},x_{\hat{i}^\prime,n})=\log\frac{\textrm{Pr}(\fat{y}_n^\shortmid|x_{\hat{i},n}=0,x_{\hat{i}^\prime,n}=1)}
{\textrm{Pr}(\fat{y}_n^\shortmid|x_{\hat{i},n}=1,x_{\hat{i}^\prime,n}=0)}\\
&&\!\!\!\!\!\!\!\!\!\!\!\!=L_1(x_{\hat{i},n},x_{\hat{i}^\prime,n})+\sum_{m\in\{m|m\neq \hat{i}, \hat{i}^\prime, \hat{s}_m^-=1/2\}}\log\frac{\textrm{Pr}(y_{m,n}|x_{\hat{i},n}=0,x_{\hat{i}^\prime,n}=1)}
{\textrm{Pr}(y_{m,n}|x_{\hat{i},n}=1,x_{\hat{i}^\prime,n}=0)}\nonumber\\
&&\!\!\!\!\!\!\!\!\!\!\!\!=L_1(x_{\hat{i},n},x_{\hat{i}^\prime,n})+\sum_{m\in\{m|m\neq \hat{i}, \hat{i}^\prime, \hat{s}_m^-=1/2\}}L_{m,n}(x_{\hat{i},n},x_{\hat{i}^\prime,n}).
\end{eqnarray}
Note that $\textrm{Pr}(y_{m,n}|x_{\hat{i},n}=0,x_{\hat{i}^\prime,n}=1)=\textrm{Pr}(y_{m,n}|x_{\hat{i},n}=1,x_{\hat{i}^\prime,n}=0)$ holds if $\hat{s}_m^-\in\{0, 1\}$ for $m\neq \hat{i}, \hat{i}^\prime$. Similarly, we have
\begin{eqnarray}
L_2(x_{m,\hat{j}},x_{m,\hat{j}^\prime})\!=\!L_1(x_{m,\hat{j}},x_{m,\hat{j}^\prime})\!+\!\!\sum_{n\in\{n|n\neq \hat{j}, \hat{j}^\prime, \hat{s}_n^\shortmid=1/2\}}\!L_{m,n}(x_{\hat{i},n},x_{\hat{i}^\prime,n}).
\end{eqnarray}

Based on $\fat{s}^\shortmid, \fat{s}^-$ and $L_2(x_{\hat{i},n}, x_{\hat{i}^\prime,n}), L_2(x_{m,\hat{j}}, x_{m,i^\prime})$, we do the final hard decision for the SF rows and columns
\begin{equation}\label{eq:est41}
(\hat{x}_{\hat{i},n},\hat{x}_{\hat{i}^\prime,n})=\begin{cases} (0, 0)&\mbox{if $s_n^\shortmid=0$}\\ (1, 1)&\mbox{if $s_n^\shortmid=1$}\\
(0, 1)&\mbox{if $s_n^\shortmid=1/2, L_2(x_{\hat{i},n}, x_{\hat{i}^\prime,n})>0$}\\
(1, 0)&\mbox{if $s_n^\shortmid=1/2, L_2(x_{\hat{i},n}, x_{\hat{i}^\prime,n})\leq0$} \end{cases}
\end{equation}
for $n =1,...,N$, and
\begin{equation}\label{eq:est42}
(\hat{x}_{m,\hat{j}},\hat{x}_{m,\hat{j}^\prime})=\begin{cases} (0, 0)&\mbox{if $s_m^-=0$}\\ (1, 1)&\mbox{if $s_m^-=1$}\\
(0, 1)&\mbox{if $s_m^-=1/2, L_2(x_{m,\hat{j}}, x_{m,\hat{j}^\prime})>0$}\\
(1, 0)&\mbox{if $s_m^-=1/2, L_2(x_{m,\hat{j}}, x_{m,\hat{j}^\prime})\leq0$} \end{cases}
\end{equation}
for $m =1,...,N$.

 \begin{figure*}[t]
\includegraphics[width=
4.5 in]{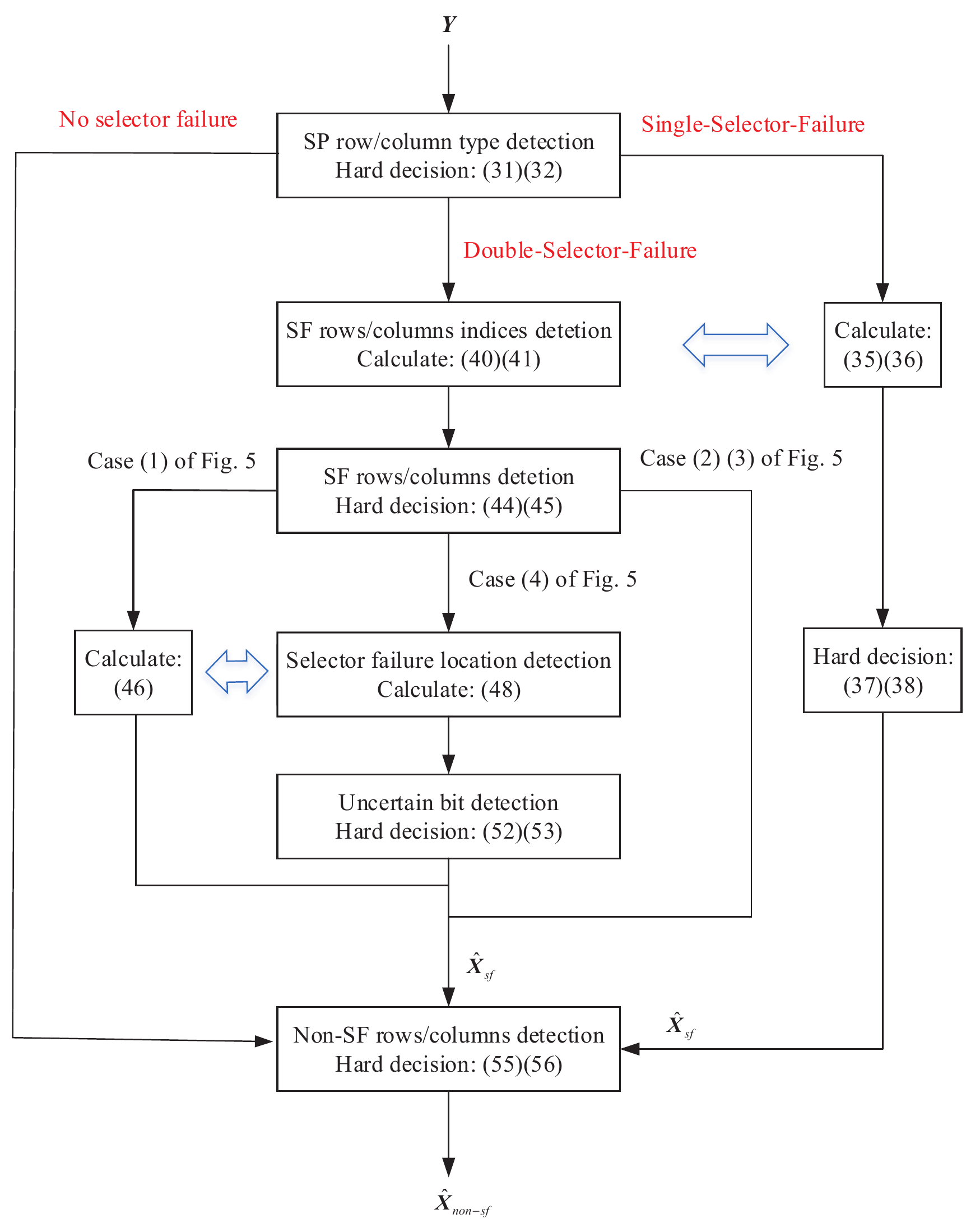}
\centering
\caption{Flow chart of the proposed data detection scheme} \label{fig:procedure}
\end{figure*}
\section{Non-SF-Row and -Column $\fat{X}_{non-sf}$ Detection}\label{sec:non-SF}
Based on $\hat{\fat{X}}_{sf}$, we can detect each entry of $\fat{X}_{non-sf}$ using MAP criteria.
For each non-SF entry $x_{m,n}$, we calculate the posteriori LLR based on $y_{m,n}$
\begin{eqnarray}
\!\!\!\!\!\!\!\!\!\!\!\!\!\!\!\!\!\!\!\!&&L(x_{m,n})
=\log\frac{\textrm{Pr}(x_{m,n}=0|y_{m,n},\hat{\fat{X}}_{sf})}{\textrm{Pr}(x_{m,n}=1|y_{m,n},\hat{\fat{X}}_{sf})}\nonumber\\
\!\!\!\!\!\!\!\!\!\!\!\!\!\!\!\!\!\!\!\!&&=\log\frac{\textrm{Pr}(x_{m,n}=0)\textrm{Pr}(y_{m,n}|x_{m,n}=0,\hat{\fat{X}}_{sf})}{\textrm{Pr}(x_{m,n}=1)\textrm{Pr}(y_{m,n}|x_{m,n}=1,\hat{\fat{X}}_{sf})}\nonumber\\
\!\!\!\!\!\!\!\!\!\!\!\!\!\!\!\!\!\!\!\!&&=\log\frac{1-q}{q}+\log\frac{\exp\left(-\frac{\left(y_{m,n}-R_0\left(\bigcup_{(i,j)\in \varphi^*}x_{i,n}x_{m,j}\right)\right)^2}{2\sigma^2}\right)}{\exp\left(-\frac{\left(y_{m,n}-R_1\right)^2}{2\sigma^2}\right)}\nonumber\\
\!\!\!\!\!\!\!\!\!\!\!\!\!\!\!\!\!\!\!\!&&=\log\frac{1-q}{q}\!+\!\frac{\left(y_{m,n}-R_1\right)^2\!-\!\left(y_{m,n}-R_0\left(\bigcup_{(i,j)\in \varphi^*}x_{i,n}x_{m,j}\right)\right)^2}{2\sigma^2}.
\end{eqnarray}
If $L(x_{m,n})>0$, $x_{m,n}=0$, otherwise, $x_{m,n}=1$. This leads to the following double thresholds decision rule.

For non-SP cell $(m, n)$ with $\bigcup_{(i,j)\in \varphi^*}x_{i,n}x_{m,j}=0$,
\begin{equation}\label{eq:nonsf}
\hat{x}_{m,n}=\begin{cases} 0&\mbox{if $y_{m,n}>\gamma$}\\ 1&\mbox{if $y_{m,n}\leq\gamma$} \end{cases}
\end{equation}
where $\gamma=\frac{\sigma^2}{R_0-R_1}\log\frac{q}{1-q}+\frac{R_0+R_1}{2}$.

\begin{figure}[t]
\includegraphics[width=
3.5 in]{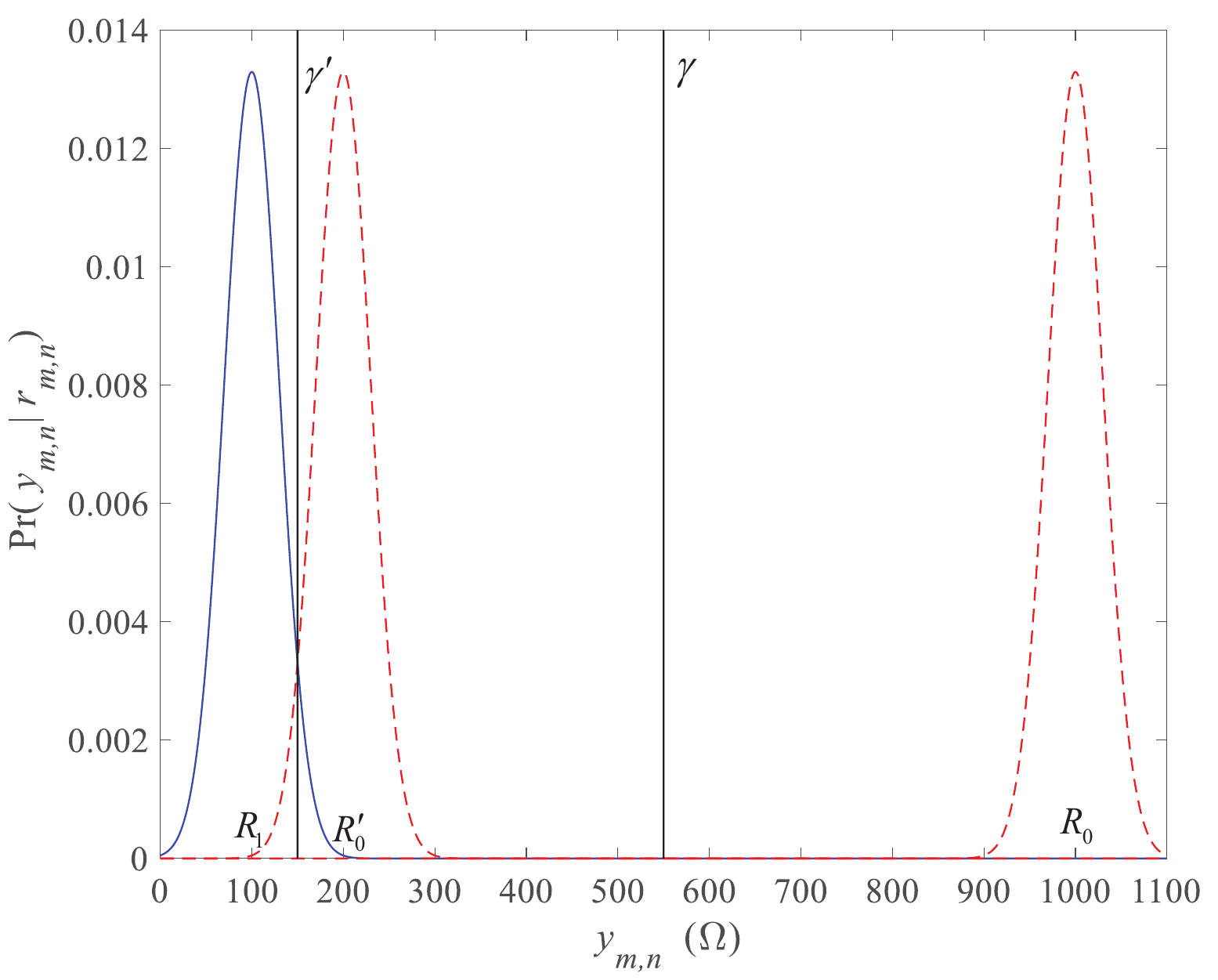}
\centering
\caption{Plots of the conditional distributions of $y_{m,n}$ for the three hypotheses: $r_{m,n}=R_1, R_0^\prime, R_0$, and the double thresholds for both non-SP cells and SP-potential cells} \label{fig:PDF_th}
\end{figure}

For SP-potential cell $(m, n)$ with $\bigcup_{(i,j)\in \varphi^*}x_{i,n}x_{m,j}=1$,
\begin{equation}\label{eq:nonsf2}
\hat{x}_{m,n}=\begin{cases} 0&\mbox{if $y_{m,n}>\gamma^\prime$}\\ 1&\mbox{if $y_{m,n}\leq\gamma^\prime$} \end{cases}
\end{equation}
where $\gamma^\prime=\frac{\sigma^2}{R_0^\prime-R_1}\log\frac{q}{1-q}+\frac{R_0^\prime+R_1}{2}$.
The double thresholds when $q=1/2$, $R_0=1000 \Omega, R_1=100 \Omega, R_s=250 \Omega$, and $\sigma=30$ are shown in Fig.~\ref{fig:PDF_th}.

A flow chart of the proposed data detection scheme is shown in Fig.~\ref{fig:procedure}. Since every execution block in the flow chart can be completed with complexity no greater than $O(N^2)$, the overall data detection complexity increases linearly with the array size.

\section{Performance Bound and Simulations}\label{sec:simulation}
In this section we propose a BER lower bound for a large memory array. Using BER simulation we show that our data detection scheme
achieves the bound.
\subsection{BER Bound}
The BER is lower bounded by when the SF rows and columns are perfectly recovered.
If the SF rows and columns are perfectly recovered, the BER of the
non-SF entries based on hard decision rules (\ref{eq:nonsf}) and (\ref{eq:nonsf2}) is
\begin{eqnarray}
P_{non-sf}=\!\!\!\!\!\!\!\!&&\textrm{Pr}\left(\bigcup_{(i,j)\in \varphi^*}x_{i,n}x_{m,j}=1\right)Q(\frac{\gamma^\prime-R_1}{\sigma})\nonumber\\
\!\!\!\!\!\!\!\!&&\ \ +\textrm{Pr}\left(\bigcup_{(i,j)\in \varphi^*}x_{i,n}x_{m,j}=0\right)Q(\frac{\gamma-R_1}{\sigma})
\end{eqnarray}
where $Q(x)=\frac{1}{2\pi}\int_{x}^\infty\exp\left(-\frac{u^2}{2}\right)du$ is the Q-function.
If there are $k$ active SFs we have $\textrm{Pr}\left(\bigcup_{(i,j)\in \varphi^*}x_{i,n}x_{m,j}=0\right)=(1-q^2)^k$ and $\textrm{Pr}\left(\bigcup_{(i,j)\in \varphi^*}x_{i,n}x_{m,j}=1\right)=1-(1-q^2)^k$ and the overall BER is $\left(1-\frac{2kN-k^2}{N^2}\right)P_{non-sf}$, where $1-\frac{2kN-k^2}{N^2}$ is the fraction of non-SF entries. If we know the a priori probability of active
SF numbers $p_k, k=0, 1, 2$, the overall BER is lower bounded by
\begin{eqnarray}
P_e\geq\!\!\!\!\!\!\!\!&&\sum_{k=0}^2p_k\left(1-\frac{2kN-k^2}{N^2}\right)\left((1-q^2)^kQ(\frac{\gamma-R_1}{\sigma})\right.\nonumber\\
\!\!\!\!\!\!\!\!&&\ \ \ \ \ \ +\left.(1-(1-q^2)^k)Q(\frac{\gamma^\prime-R_1}{\sigma})\right).\label{eq:bound1}
\end{eqnarray}
We believe that when the array size is $N\rightarrow\infty$, the SF rows and columns can always be detected accurately and hence the equality in (\ref{eq:bound1}) holds.
Therefore, when $N\rightarrow\infty$, $\frac{2kN-k^2}{N^2}\rightarrow0$, and we have asymptotical BER bound:
\begin{eqnarray}
P_e^{N\rightarrow\infty}= (1-P_{sp})Q(\frac{\gamma-R_1}{\sigma})+P_{sp}Q(\frac{\gamma^\prime-R_1}{\sigma})\label{eq:bound}
\end{eqnarray}
where $P_{sp}=1-\sum_{k=0}^2p_k(1-q^2)^k$ is the occurrence probability of an SP-potential cell. Note that the bound of (\ref{eq:bound}) is the lowest BER that can be achieved by an optimal data detection scheme.

 \begin{figure}[t]
\includegraphics[width=
3.5 in]{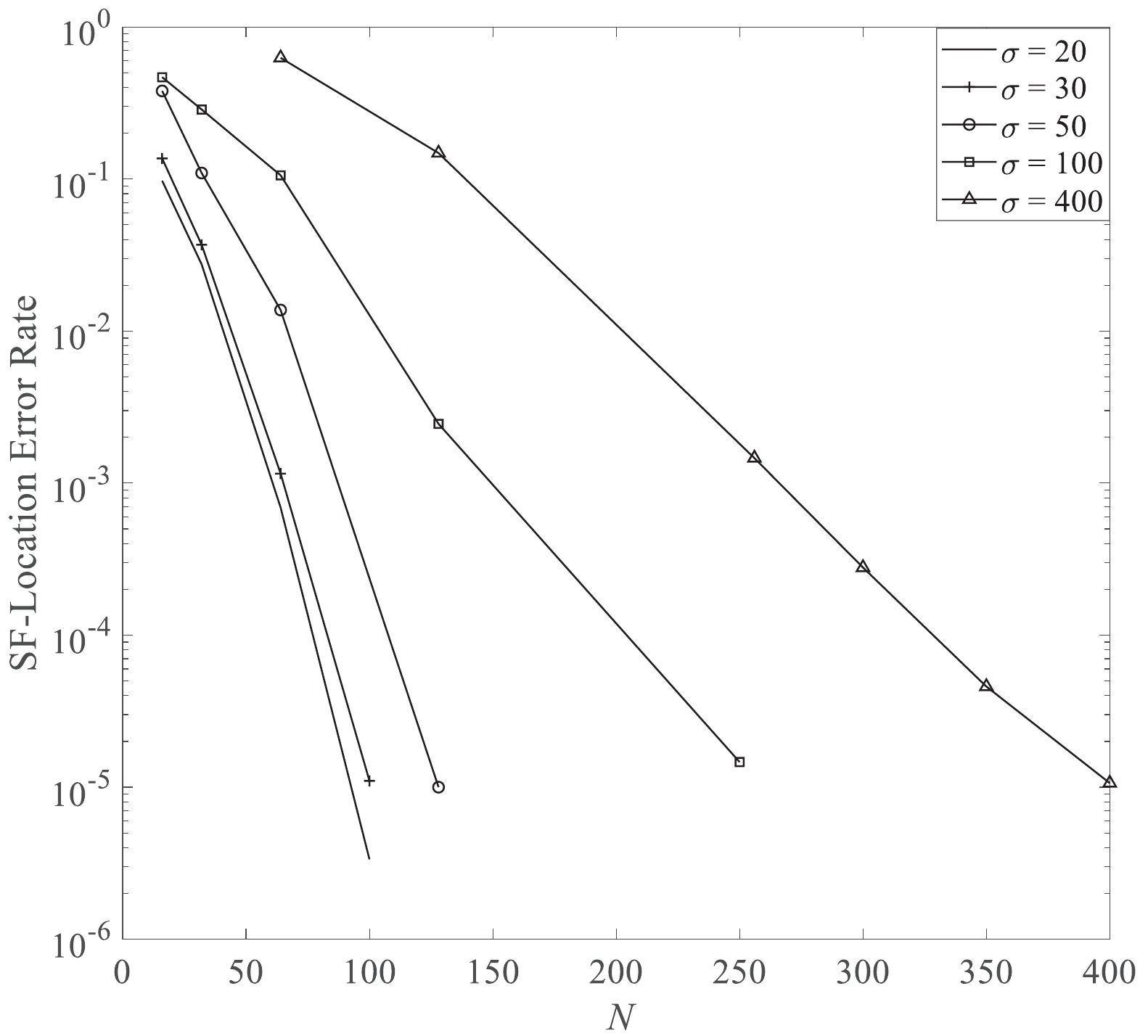}
\centering
\caption{Detection error rate of SF locations when the SF probability distribution is $\fat{p}_a=(0.5, 0.4, 0.1)$} \label{fig:SF-loc-error}
\end{figure}

 \begin{figure}[t]
\includegraphics[width=
3.5 in]{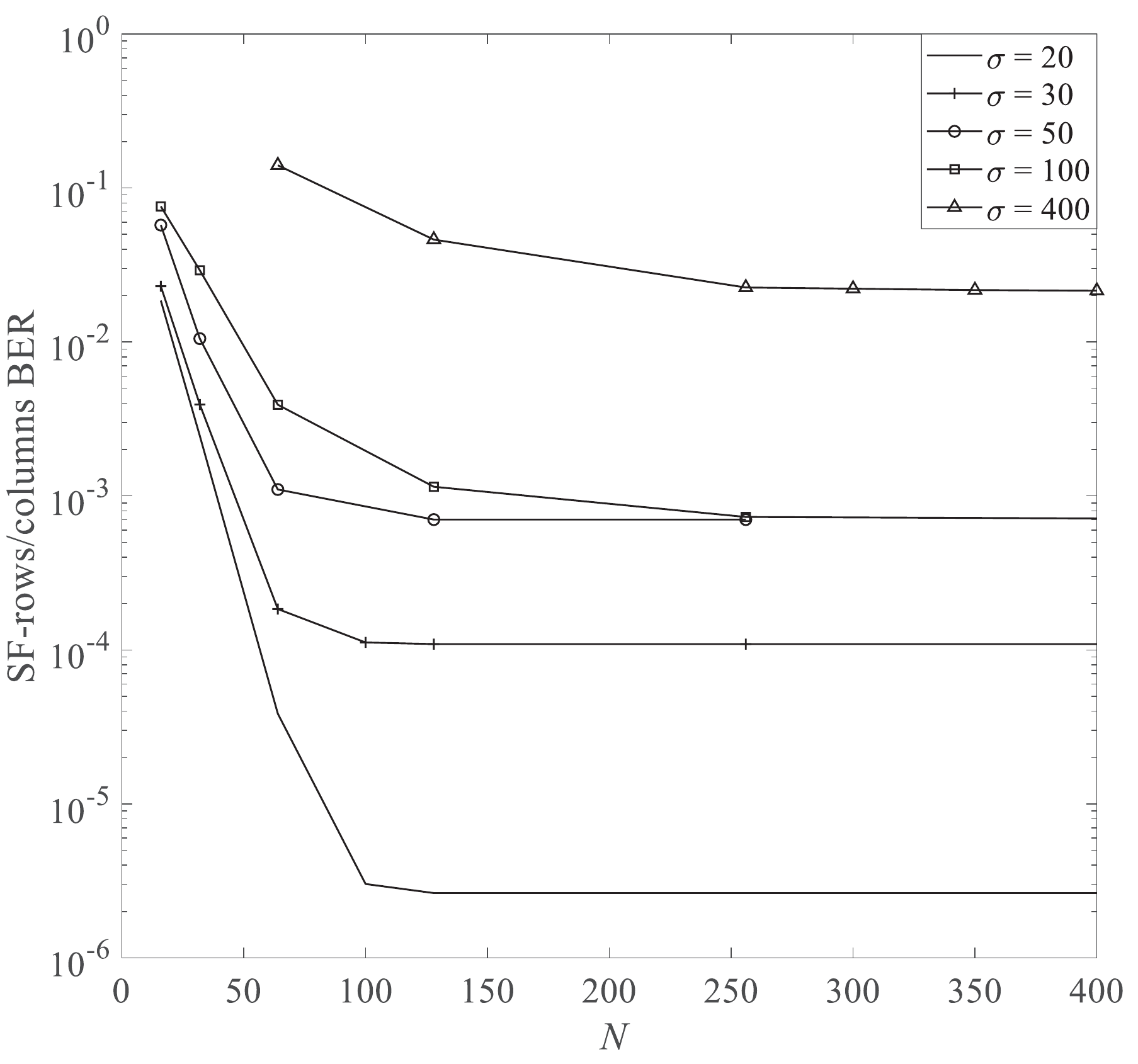}
\centering
\caption{Detection BER of SF rows and columns when the SF probability distribution is $\fat{p}_a=(0.5, 0.4, 0.1)$} \label{fig:SF-BER}
\end{figure}
\subsection{Simulation Results}
In this section, we provide some simulation results for the performance of our detection scheme and compare our scheme with both the previous scheme and the BER bound. We use $q=1/2$, $R_0=1000 \Omega, R_1=100 \Omega$, and $R_s=250 \Omega$ as our simulation parameters.

Figures~\ref{fig:SF-loc-error} and \ref{fig:SF-BER} show the detection error rate of SF locations and the detection BER of SF rows and columns when the SF probability distribution is set as $\fat{p}_a=(p_0, p_1, p_2)=(0.5, 0.4, 0.1)$. It shows that as the array size $N$ increases, the detection for the SF locations is very accurate even when $\sigma=400$. The detection BER for the SF rows and columns decreases as the array size increases, while if the array size is sufficiently large, $(>150)$, it approaches a constant value that is determined by the noise level.

\begin{figure}[t]
\includegraphics[width=
3.5 in]{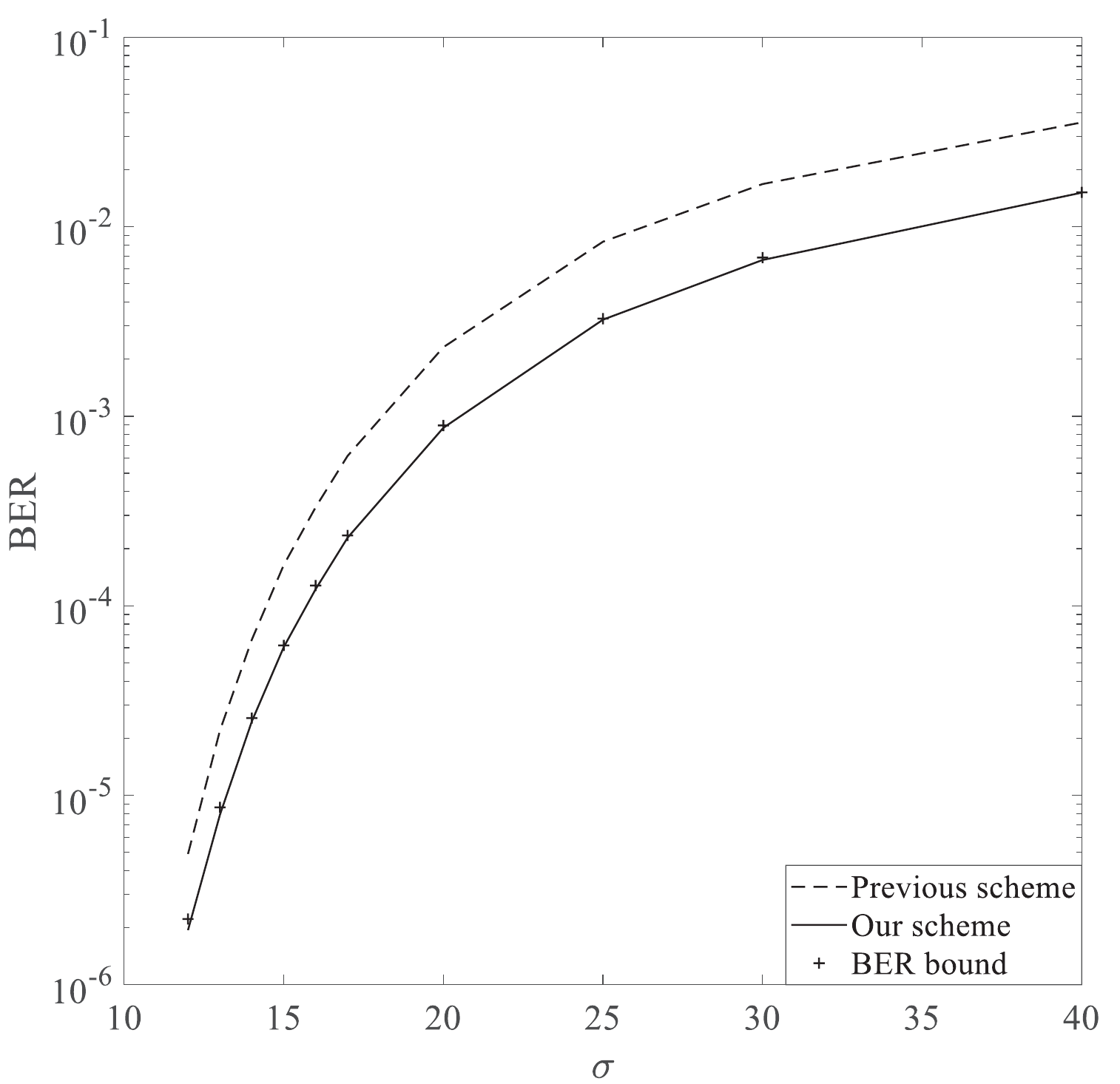}
\centering
\caption{Data detection BER when $N=128$ and the SF probability distribution is $\fat{p}_a=(0.5, 0.4, 0.1)$. The BER of the previous scheme shows the results for the detection scheme proposed in Ben-Hur and Cassuto \cite{Ben} (the single-threshold scheme used in Chen et al. \cite{CZH}).} \label{fig:BER128a}
\end{figure}

\begin{figure}[t]
\includegraphics[width=
3.5 in]{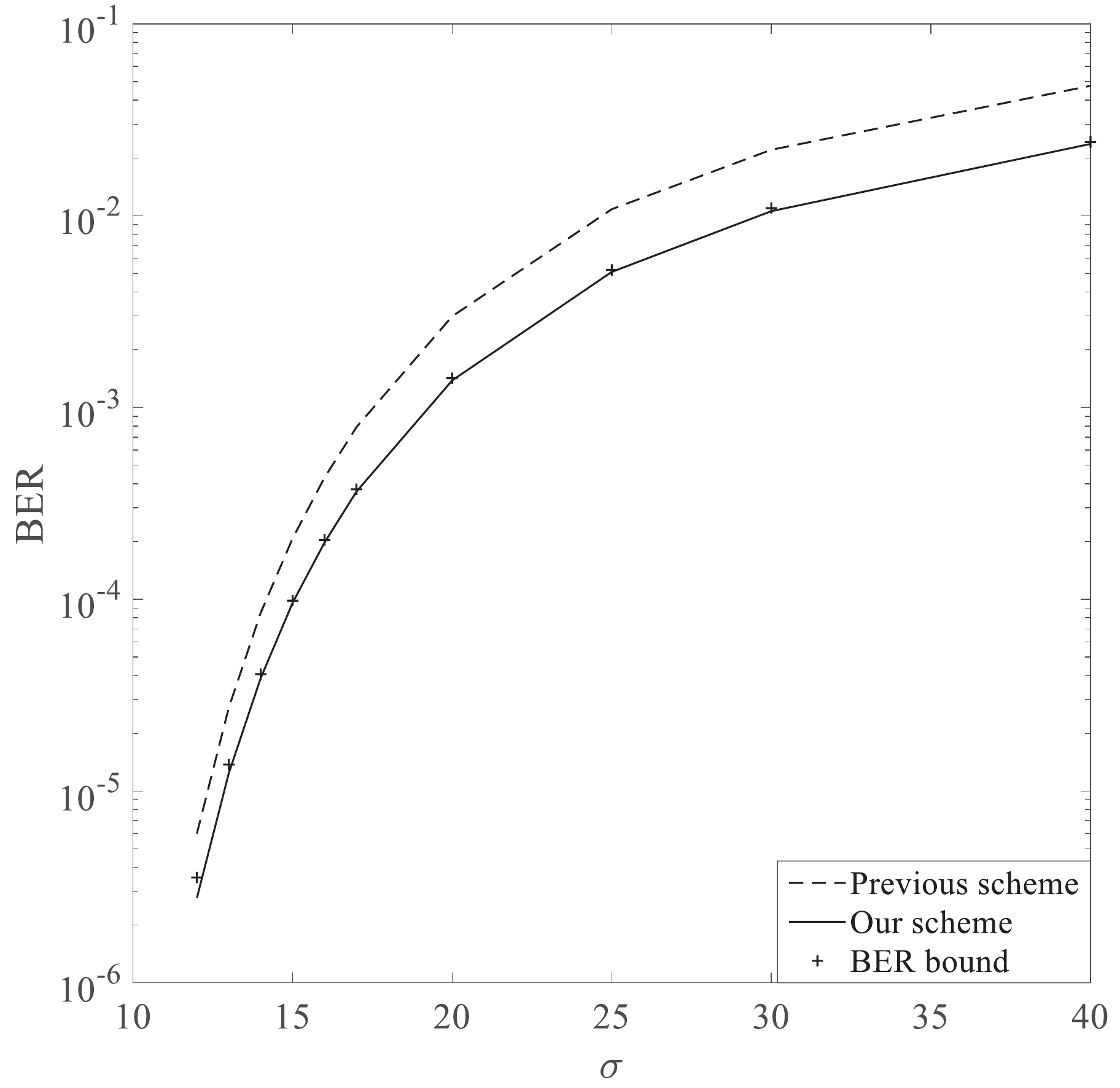}
\centering
\caption{Data detection BER when $N=128$ and the SF probability distribution is $\fat{p}_b=(1/3, 1/3, 1/3)$. The BER of the previous scheme shows the result for the detection scheme proposed in Ben-Hur and Cassuto \cite{Ben} (the single-threshold scheme used in Chen et al. \cite{CZH}).} \label{fig:BER128b}
\end{figure}

Figures~\ref{fig:BER128a} and \ref{fig:BER128b} provide the overall detection BER for the data array involving both of the SF rows and columns and non-SF rows and columns when $N=128$ and the SF probability distributions are set as $\fat{p}_a$ and $\fat{p}_b=(1/3, 1/3, 1/3)$, respectively. As shown in both figures, our data detection scheme outperformed the scheme proposed in Ben-Hur and Cassuto \cite{Ben} (the single-threshold scheme used in Chen et al. \cite{CZH}). The performance gain is more obvious when the noise level is large. The performance at high noise levels dominates the BER performance when error correction codes are used in the system. The reason for the performance gain is that our scheme accurately detects the SF locations and
SPs while the previous work of Ben-Hur and Cassuto \cite{Ben} and Chen et al. \cite{CZH} detected the data by regarding the SP interference as noise. Our detection scheme can approach the BER bound (\ref{eq:bound}), which implies its optimality.


\section{Conclusion} \label{sec:conclusion}
 We proposed a near-optimal data detection scheme for a ReRAM system. This scheme can take advantage of inter-cell correlation and jointly recover both of the data and SP interference. It can approach the performance bound of the optimal detection scheme with only linear operation complexity.

In order to guarantee a perfect recovery of the SF and SP statuses, we limited the number of SFs in an array to be fewer than three (this is also the most likely case) and assumed the array size was large. However, many of the ideas and the data detection principles proposed in this work can be used to deal with situations in which there are more than two SFs and the array size is not large, even though in these cases, perfect recoveries of the SF and SP status might be difficult. For example, we can develop an adaptive thresholding strategy based only on the SP-row or -column type or based on the soft estimation of the SF rows or columns rather than fully recovering the SF rows or columns.


\begin{thebibliography}{1}

\bibitem{Strukov} D. B. Strukov, G. S. Snider, D. R. Stewart, and R. S. Williams, ``The
missing memristor found," Nature, vol. 453, no. 7191, p. 80, 2008.

\bibitem{Zidan} M. A. Zidan, H. A. H. Fahmy, M. M. Hussain, and K. N. Salama,
``Memristor-based memory: The sneak paths problem and solutions,"
Microelectron. J., vol. 44, no. 2, pp. 176--183, 2013.

\bibitem{Yuval}
Y. Cassuto, S. Kvatinsky, and E. Yaakobi, ``Information-theoretic sneakpath mitigation in memristor crossbar arrays," \emph{IEEE Trans. Inf. Theory.}, vol. 62, no. 9, pp. 4801--4813, Sep. 2016.


\bibitem{Ben} Y. Ben-Hur and Y. Cassuto, ``Detection and coding schemes for sneakpath interference in resistive memory arrays," \emph{IEEE Trans.
Commun.}, vol. 67, no. 6, pp. 3821--3833, Feb. 2019.

\bibitem{CZH} Z. Chen, C. Schoeny, and L. Dolecek, ``Pilot assisted adaptive thresholding for
sneak-path mitigation in resistive memories with
failed selection devices," \emph{IEEE Trans.
Commun.}, vol. 68, no. 1, pp. 66--81, Jan. 2020.

\bibitem{SongArchive} G. Song, K. Cai, X. Zhong, J. Yu, and J. Cheng, ``Performance limit and coding schemes for resistive random-access memory channels," \emph{arXiv:2005.02601 [cs.IT]}

\bibitem{Yu2012} S. Yu, X. Guan and H.-S. P. Wong, ``On the switching parameter variation of metal oxide RRAM -- Part II: model corroboration and device design strategy," \emph{IEEE Trans. Electron Devices}, vol. 59, no. 4, pp. 1183--1188, Apr. 2012.

\bibitem{Wong2012} H.-S. P. Wong et al., ``Metal-oxide RRAM," in \emph{Proc. IEEE}, vol. 100, no. 6, pp. 1951--1970, Jun. 2012.



\end{thebibliography}
\end{document}